\pdfoutput=1
\documentclass[11pt,a4paper]{article}
\usepackage{jheppub}
\allowdisplaybreaks[1]
\usepackage{bm}
\usepackage{subfig}

\begin{document}

\title{Hairy black holes in AdS with Robin boundary conditions}
\author[1]{Tomohiro~Harada,}
\author[1]{Takaaki~Ishii,}
\author[2,3]{Takuya~Katagiri,}
\author[4]{and Norihiro~Tanahashi}
\affiliation[1]{Department of Physics, Rikkyo University, Nishi-Ikebukuro, Toshima, Tokyo 171-8501, Japan}
\affiliation[2]{Astronomical Institute, Graduate School of Science, Tohoku University, Aoba, Sendai 980-8578, Japan}
\affiliation[3]{Niels Bohr International Academy, Niels Bohr Institute, Blegdamsvej 17, 2100 Copenhagen, Denmark}
\affiliation[4]{Department of Physics, Chuo University, Kasuga, Bunkyo, Tokyo 112-8551, Japan}
\emailAdd{harada@rikkyo.ac.jp}
\emailAdd{ishiitk@rikkyo.ac.jp}
\emailAdd{takuya.katagiri@astr.tohoku.ac.jp}
\emailAdd{tanahashi@phys.chuo-u.ac.jp}

\abstract{We study hairy black holes in Einstein-Maxwell-complex scalar theory in four-dimensional asymptotically global anti-de Sitter (AdS) spacetime when the Robin boundary conditions are imposed on the scalar field. This setup is dual to the double trace deformation of strongly interacting field theory on $R \times S^2$ by charged scalar operators. We identify the instability of the Reissner-Nordstr\"{o}m-AdS (RNAdS) black holes under the Robin boundary conditions and construct backreacted geometries branching at the onset of the instability. Also considering associated horizonless geometries called boson stars, we obtain phase diagrams with fairly rich structure in the grand canonical ensemble depending on the boundary condition parameter or the deformation parameter, where phase transition occurs between thermal AdS, RNAdS, charged boson stars, and hairy black holes.}
\preprint{RUP-23-6}

\maketitle

\section{Introduction}

Asymptotically anti-de Sitter (AdS) spacetime offers diverse gravitational dynamics. In contrast to asymptotically flat spacetime, black hole geometry can be considered in the canonical ensemble, where asymptotically global AdS experiences the first order phase transition between horizonless and black hole spacetimes \cite{Hawking:1982dh,Witten:1998zw}. Through the AdS/CFT duality~\cite{Maldacena:1997re,Witten:1998qj,Gubser:1998bc}, it is interpreted as the confinement/deconfinement phase transition in strongly coupled Yang-Mills theory. When the gravitational theory has $U(1)$ gauge field and charged scalar field, the spontaneous breaking of the gauge symmetry is discussed as the appearance of the superfluid/superconducting phase \cite{Gubser:2008px,Hartnoll:2008vx,Hartnoll:2008kx}.

Aforementioned phenomena are often considered with the Dirichlet boundary conditions imposed on the asymptotic behavior of the scalar field at the AdS boundary.
However, general conditions known as the Robin boundary conditions (also called mixed boundary conditions) are allowed \cite{doi:10.1063/1.524403,Ishibashi:2003jd,Ishibashi:2004wx,Henneaux:2004zi,Henneaux:2006hk} if the field in AdS has a mass close to the Breitenlohner-Freedman bound \cite{Breitenlohner:1982bm,Breitenlohner:1982jf}. When the parameter for the Robin boundary conditions exceeds a critical value
and the deviation from the Dirichlet boundary condition becomes sufficiently large, 
the AdS spacetime becomes unstable \cite{Ishibashi:2004wx}. The Robin (or mixed) boundary conditions are related to multitrace deformation in the dual field theory in the AdS/CFT interpretation~\cite{Klebanov:1999tb,Witten:2001ua,Berkooz:2002ug}. Not only for scalar field considered in these literature, but also the Robin boundary conditions can be imposed for vector field and discussed in the context of introducing dynamical gauge field on the AdS boundary \cite{Ishibashi:2004wx,Witten:2003ya,Marolf:2006nd}. Robin boundary conditions have also been considered for metric field so as to promote the boundary metric dynamical \cite{Compere:2008us}.

In~\cite{Katagiri:2020mvm}, two of the authors studied the linear mode stability of the four-dimensional Reissner-Nordstr\"{o}m AdS (RNAdS) spacetime with global AdS asymptotics for neutral and charged complex scalar field perturbations with Robin boundary conditions.\footnote{There is a recent work on the quasinormal mode spectrum of a scalar field with the Robin boundary conditions in Schwarzschild AdS$_4$ spacetime \cite{Kinoshita:2023iad}. See also superradiance in BTZ black holes with the Robin boundary conditions \cite{Dappiaggi:2017pbe}.} The neutral field shows an instability for the Robin boundary conditions with parameters greater than a critical value. 
The charged scalar field suffers another type of instability due to
the electromagnetic interaction with the black hole, which is known as superradiance~\cite{PhysRevD.7.949,Hawking:1999dp,Uchikata:2011zz,Green:2015kur,Dias:2016pma}.\footnote{Instability of RNAdS can be associated with the violation of near horizon AdS$_2$ BF bound, but it is a necessary condition. For charged scalar, superradiance occurs regardless \cite{Ishii:2022lwc}, so here we simply describe the cause of this charged instability as superradiance.} With the imposition of the Robin boundary conditions, superradiance and the boundary contribution interplay with each other, potentially enhancing the instability caused by the superradiance depending on the parameters of the scalar field and the background spacetime. It was argued in~\cite{Katagiri:2020mvm} that the instability would change the RNAdS to charged hairy black hole solutions with a nontrivial scalar field satisfying the Robin boundary conditions, which are a candidate for the final fate of the instability.
First studied for neutral scalar, the presence of hairy solutions with the Robin boundary conditions has been known; see \cite{Hertog:2004dr,Martinez:2004nb,Hertog:2004ns} for early works. Motivated by \cite{Katagiri:2020mvm}, we study charged hairy solutions in four dimensional global AdS spacetime in detail.

In this paper, we study hairy black holes that branch at the onset of instability of the charged scalar field with the Robin boundary conditions on the four-dimensional RNAdS,
and obtain results that agree with the expectation of \cite{Katagiri:2020mvm} explained above.
Following~\cite{Hartnoll:2008vx,Hartnoll:2008kx}, hairy black holes have been widely studied in Einstein-Maxwell-complex scalar theory in asymptotically AdS spacetime, in both Poincar\'{e} and global AdS spacetimes and in various dimensions.
In studies of this sort, the Dirichlet (and Neumann) boundary conditions are often considered. For example,
the phase diagram in asymptotically global AdS$_4$
in the grand canonical ensemble was explored in \cite{Basu:2016mol}.\footnote{Hairy black holes have been also considered in global AdS$_5$ \cite{Maeda:2010hf,Basu:2010uz,Bhattacharyya:2010yg,Dias:2011tj}. See also \cite{Mahapatra:2020wym,Priyadarshinee:2021rch}.}
In this paper, we conduct a comprehensive study on the phase structures realized under the Robin boundary conditions in the grand canonical ensemble.
Within the four dimensional global AdS spacetime, charged scalar solitons (boson stars) and hairy black holes in setups including the same model as ours have been considered in \cite{Gentle:2011kv}.\footnote{See also prior works in three dimensions \cite{Henneaux:2002wm,Correa:2010hf}.
See also a recent study of boson stars of mixed boundary conditions deformation~\cite{Liu:2020uaz,Guo:2020bqz}
motivated by the analysis on the large charge limit in CFT~\cite{Hellerman:2015nra}.}
Our work may be viewed as a generalization of this work, clarifying the full phase structure of such solutions under the Robin boundary conditions.

This paper is organized as follows. In section~\ref{sec:setup}, we prepare the setup for constructing boson stars and hairy black holes with the Robin boundary conditions. In particular, we study the onset of instability of the four dimensional RNAdS spacetime with respect to the charged scalar field perturbations with the Robin boundary conditions. In section~\ref{sec:results}, we show results of the phase diagram for our setup under the Robin boundary conditions.
Section~\ref{sec:conclusion} concludes the paper.
In appendix~\ref{sec:holoreno}, we summarize holographic renormalization for the Robin boundary conditions.
In appendix~\ref{sec:1st}, we discuss the first law of thermodynamics.
In appendix~\ref{sec:micro}, we comment on entropies in microcanonical ensemble. 

\section{Setup}
\label{sec:setup}

\subsection{Reissner-Nordstr\"{o}m AdS black hole}

We consider Einstein-Maxwell-complex scalar theory in four-dimensional asymptotically global AdS spacetime. The action is
\begin{equation}
S = \frac{1}{8 \pi G_N} \int \mathrm{d}^4 x \sqrt{-g} \left(\frac{1}{2}\left( R - 2 \Lambda \right) - \frac{1}{4} F_{\mu\nu} F^{\mu\nu} - |D \phi|^2 - m^2 |\phi|^2 \right),
\label{EMS_action}
\end{equation}
where $F_{\mu\nu}=\partial_\mu A_\nu-\partial_\nu A_\mu$, and $D_\mu \phi = \partial_\mu \phi - i q A_\mu \phi$. The gauge coupling constant is written by $q$. We use units in which $\Lambda = - 3$ so that the AdS radius can be set to unity. The mass of the scalar is related to the conformal dimension of the scalar operator in the dual field theory as $m^2 =\Delta(\Delta-3)$. We set $m^2=-2$ in this paper. Then, this equation is solved by $\Delta=1,2$. The equations of motion are
\begin{equation}
G_{\mu\nu} + \Lambda g_{\mu\nu} = T_{\mu\nu}, \quad
\nabla_\mu F^{\mu\nu} = J^{\nu}, \quad
\left( D_\mu D^\mu - m^2 \right) \phi =0,
\label{EoMs}
\end{equation}
where
\begin{equation}
\begin{split}
T_{\mu\nu} &= F_{\mu\lambda}F_{\nu}{}^{\lambda} + (D_\mu \phi)^* D_\nu\phi + (D_\nu \phi)^* D_\mu\phi + g_{\mu\nu} \mathcal{L}, \\
\mathcal{L} &= - \frac{1}{4} F_{\mu\nu} F^{\mu\nu} - |D \phi|^2 - m^2 |\phi|^2, \\
J^{\mu} &= 2 q^2 |\phi|^2 A^\mu + i q (\phi^* \partial^\mu \phi - \phi \partial^\mu \phi^*).
\end{split}
\end{equation}

We study spherically symmetric static solutions in the spherical AdS boundary. The ansatz can be given by
\begin{align}
\mathrm{d} s^2 &= - \left( 1+ r^2 \right) f(r) e^{-\chi(r)} \mathrm{d}t^2 + \frac{\mathrm{d}r^2}{\left( 1+ r^2 \right)f(r)} + r^2 \mathrm{d}\Omega_2^2, \label{metric_ansatz} \\
A &= A_t(r) \mathrm{d} t, \qquad \phi=\phi(r). \label{At_ansatz}
\end{align}
The conformal boundary of the AdS is $R \times S^2$ and located at $r=\infty$. When $f(r)=1$ and $\chi(r)=0$ (as well as $A=\phi=0$), the empty AdS is obtained.
For horizonless geometries, $r=0$ is the center of the AdS.

The RNAdS black hole is given by
\begin{equation}
\begin{split}
\left( 1+ r^2 \right) f(r) &= 1 + r^2 - \left( 1 + r_h^2 + \frac{Q^2}{2 r_h^2} \right) \frac{r_h}{r} + \frac{Q^2}{2 r^2}, \\
A_t(r) &= \mu - \frac{Q}{r}, \quad \chi(r)=\phi(r)=0,
\end{split}
\label{RNAdSBG}
\end{equation}
where $r_h$ denotes the location of the outermost horizon, satisfying $f(r_h)=0$, and $Q$ is the charge of the black hole per solid angle. The total charge is given by $\mathcal{Q} = 4 \pi Q$. We choose the gauge as $A_t(r_h) = 0$, and then we obtain $\mu = Q/r_h$, where $\mu$ is identified as the chemical potential of the gauge field. For the diagonal metric \eqref{metric_ansatz}, the Hawking temperature and the Bekenstein-Hawking entropy are given by
\begin{align}
T_\mathrm{H} &= \frac{1}{4\pi}(1+r_h^2)f'(r_h)e^{-\chi(r_h)/2}, \label{T_Hawking} \\
S_\mathrm{BH} &= \frac{8 \pi^2 r_h^2}{8 \pi G_N}. \label{S_RN}
\end{align}

For the RNAdS, the temperature is
\begin{equation}
T_\mathrm{H}=\frac{2(1+3 r_h^2)-\mu^2}{8 \pi r_h}.
\label{T_RN}
\end{equation}
If $\mu^2<2$, the temperature has the minimum $T_\mathrm{H}=T_0$, when
\begin{equation}
r_h = \frac{\sqrt{2- \mu^2}}{\sqrt{6}} \equiv r_0, \quad
T_0 = \frac{\sqrt{3(2- \mu^2)}}{2 \sqrt{2} \, \pi} = \frac{3 r_0}{2 \pi}.
\label{RN_minT}
\end{equation}
Black holes with $r_h>r_0$ are called large black holes, while those with $r_h<r_0$ are small. In the grand canonical ensemble, the first order transition known as the Hawking-Page transition occurs between the RNAdS and AdS when \cite{Hawking:1982dh,Witten:1998zw,Chamblin:1999hg}
\begin{equation}
r_h = \frac{\sqrt{2-\mu^2}}{\sqrt{2}} \equiv r_\mathrm{HP}, \quad
T_\mathrm{HP} = \frac{\sqrt{2-\mu^2}}{\sqrt{2} \, \pi} = \frac{r_\mathrm{HP}}{\pi}.
\label{RN_HP}
\end{equation}
The horizon radius, or temperature, of this transition can be determined by comparing grand potentials between Euclidean RNAdS \eqref{F_RN} and thermal AdS geometries. The solution with the lower grand potential is identified to be realized physically.
The RNAdS is favored over the thermal AdS in $r>r_\mathrm{HP}$, and vice versa. Note that $r_\mathrm{HP}>r_0$.
In the grand canonical ensemble, the phase in $T > T_\mathrm{H}$ is the {\it RNAdS black hole phase}. The phase in $T < T_\mathrm{H}$ corresponds to horizonless AdS geometry, which we refer to as the {\it thermal AdS phase.}
If $\mu^2>2$, the temperature~\eqref{T_RN} becomes zero when
\begin{equation}
\mu = \sqrt{2(1+3 r_h^2)} \equiv \mu_\mathrm{ext}.
\label{mu_ext}
\end{equation}
This is when the RNAdS black hole becomes extremal. For fixed $r_h$, the range of $\mu$ is bounded from above as $\mu \le \mu_\mathrm{ext}$. Note that both $T_\mathrm{HP}$ and $T_0$ become zero at the borderline value $\mu^2=2$. 
Therefore, for $\mu^{2} >2$, the Hawking-Page transition does not appear in the phase diagram, and the zero temperature geometry is the extremal RNAdS.

To solve the equations of motion, it is convenient to use the $z$-coordinate defined by $z \equiv 1/r$. In this coordinate, the AdS boundary is located at $z=0$.  By the coordinate change, the metric \eqref{metric_ansatz} can be rewritten as
\begin{equation}
\mathrm{d} s^2 = \frac{1}{z^2} \left[ - \left( 1+ z^2 \right) f(z) e^{-\chi(z)} \mathrm{d}t^2 + \frac{\mathrm{d}z^2}{\left( 1+z^2 \right)f(z)} + \mathrm{d}\Omega_2^2 \right].
\label{metric_ansatz_z}
\end{equation}
The RNAdS black hole solution \eqref{RNAdSBG} becomes
\begin{equation}
\left( 1+ z^2 \right) f(z) = 1 + z^2 - \left( 1 + z_h^2 + \frac{Q^2 z_h^4}{2} \right) \frac{z^3}{z_h^3} + \frac{Q^2 z^4}{2}, \quad A_t = \mu \left(1 - \frac{z}{z_h} \right),
\label{RNAdSBG_z}
\end{equation}
where $z_h\equiv1/r_h$.

\subsection{Instability of RNAdS}
\label{sec:instability-of-RNAdS}

We consider spherically symmetric scalar field perturbations of the RNAdS, $\phi(z)e^{-i\omega t}$. The Klein-Gordon equation takes the form
\begin{equation}
\phi'' + \left(\frac{F'}{F} - \frac{2}{z} \right) \phi' - \left( \frac{m^2}{z^2 F} - \frac{(\omega+q A_t)^2}{F^2} \right) \phi = 0,
\label{pert_KGeq_omega}
\end{equation}
where $' \equiv \partial_z$,  $F=(1+z^2) f$, $m^2=-2$, and $f,A_t$ are given by the RNAdS background~\eqref{RNAdSBG_z}. Because of the presence of the horizon, the frequency $\omega$ is complex in general. The imaginary part of the frequency is negative $\mathrm{Im}\, \omega <0$ if the perturbation is stable, and positive~$\mathrm{Im}\, \omega>0$ if instability is induced in the RNAdS background. The border $\mathrm{Im}\, \omega=0$ is the onset of instability. In the gauge we use, $A_t(z_h) = 0$, both the real and imaginary parts of $\omega$ become zero simultaneously at the onset of instability, $\mathrm{Re}\,\omega=\mathrm{Im}\,\omega=0$.\footnote{Another gauge is often used that the gauge field vanishes asymptotically while it is nonzero on the horizon, ${A}_t \to 0 \ (z \to 0)$ and ${A}_t(z_h) \neq 0$. In that gauge, the perturbation $\phi = e^{-i \omega t} \phi(z)$ has a nonzero real part $\mathrm{Re}\,\omega \neq 0$ at the onset of instability $\mathrm{Im}\,\omega=0$~\cite{Katagiri:2020mvm}. However, this frequency-dependence in the real part can be absorbed by the gauge choice. In this paper, we use a gauge where $\mathrm{Re}\,\omega=\mathrm{Im}\,\omega=0$ at the onset of instability.} This means that, to search the onset of instability of $\phi$, it is sufficient to assume the static perturbation~$\phi(z)$ and find nontrivial normal modes.

At the onset of instability $\omega=0$, \eqref{pert_KGeq_omega} is reduced to a static perturbation equation,
\begin{equation}
\phi'' + \left(\frac{F'}{F} - \frac{2}{z} \right) \phi' - \left( \frac{m^2}{z^2 F} - \frac{q^2 A_t^2}{F^2} \right) \phi = 0,
\label{pert_KGeq}
\end{equation}
which depends on three parameters $(r_h, \mu,q)$ for given $m$. For the onset of instability, we search normal mode solutions to \eqref{pert_KGeq} when boundary conditions are imposed at $z=0$ and $z=z_h$. On the horizon $z=z_h$, we impose regularity (which used to be the ingoing wave boundary condition if $\omega \neq 0$, away from the onset of instability).

We impose Robin boundary conditions at the AdS boundary $z=0$. For $m^2 = - 2$, the asymptotic behavior of $\phi$ in $z \to 0$ takes the form
\begin{equation}
\phi = \phi_1 z + \phi_2 z^2 + \cdots,
\label{phi_boundary_series}
\end{equation}
where $\phi_1$ and $\phi_2$ are integration constants. Because the scalar mass is in the range $-9/4 \le m^2 \le -5/4$, both asymptotic behaviors $\phi \sim z$ and $\phi \sim z^2$ are normalizable \cite{Balasubramanian:1998sn}. This means that both coefficients 
$\phi_1$ and $\phi_2$ can be nonzero for normalizable normal modes. The boundary conditions with $\phi_1=0$ and $\phi_2 \neq 0$ are called Dirichlet, and those with $\phi_1\neq 0$ and $\phi_2 = 0$ are Neumann. The case with general values of $\phi_1 \neq 0$ and $\phi_2 \neq 0$ is called the Robin boundary conditions. The Robin boundary conditions can be specified by a parameter $\zeta$ defined by
\begin{equation}
\cot \zeta = \frac{\phi_2}{\phi_1}.
\label{Robin_BC}
\end{equation}
We choose the domain of $\zeta$ to be periodic in $0 \le \zeta < \pi$. The points $\zeta=0$ and $\zeta=\pi/2$ correspond to the Dirichlet and Neumann boundary conditions, respectively.

Under the Robin boundary conditions, we search the onset of instability for the scalar field perturbation in the four-dimensional parameter space $(\zeta, r_h, \mu, q)$. Technically, for a set of three parameters $(r_h, \mu, q)$, we integrate the perturbation equation \eqref{pert_KGeq} from the horizon to the AdS boundary and read off the asymptotic coefficients $\phi_1$ and $\phi_2$ in \eqref{phi_boundary_series}, from which $\zeta$ can be obtained.
This procedure gives a location of the onset of instability in the $(\zeta, r_h, \mu, q)$ parameter space. 
Iterating this procedure while varying the values for the three parameters $(r_h, \mu, q)$, we obtain a relation among the four parameters
$(\zeta, r_h, \mu, q)$.
Thus, for instance, fixing $(r_h,q)$, we obtain the onset of instability is given as a curve in $(\mu,\zeta)$ plane.

In the horizonless limit $r_h=0$, the perturbation equation \eqref{pert_KGeq} can be solved analytically. The background is the global AdS $f=1$ with a constant gauge field $A_t=\mu$. The perturbation equation \eqref{pert_KGeq} then becomes
\begin{equation}
\phi'' - \frac{2}{z(1+z^2)} \phi' - \left(\frac{m^2}{z^2 (1+z^2)} - \frac{\mu^2q^2}{(1+z^2)^2} \right) \phi = 0.
\label{pert_KGeq_horizonless}
\end{equation}
When the horizon is absent, we impose $\phi'(z)|_{z=\infty}=0$ at the center of the AdS. With this boundary condition and $m^2 =-2$, \eqref{pert_KGeq_horizonless} is solved by
\begin{equation}
\phi(z) = \frac{z}{\mu q} \sin \left( \mu q \cot^{-1} z \right),
\end{equation}
which is normalized as $\phi(z)|_{z=\infty}=1$. Expanding this around $z=0$, we find~\cite{Katagiri:2020mvm,Ishibashi:2004wx}
\begin{equation}
\cot \zeta = -\frac{\mu q}{\tan(\pi \mu q/2)}.
\label{eq:cotzeta}
\end{equation}
For $r_h=0$, $\mu$ and $q$ always show up in a pair $\mu q$. The set of the parameters $(\zeta,\mu, q)$ satisfying the above relation gives a normal mode in the global AdS. While the global AdS is stable against linear perturbations, nontrivial scalar solutions branch from the AdS at the normal modes. For this reason, with a slight abuse of terminology, we also refer to the location of the AdS normal modes as the onset of instability.

\begin{figure}[t]
\centering
\subfloat[][$r_h=0$]{\includegraphics[height=5cm]{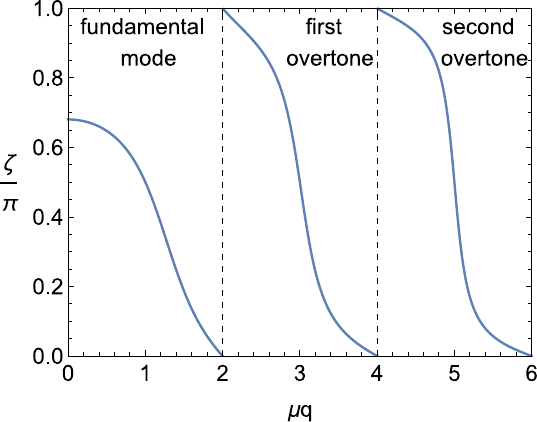}\label{fig:onset_rh0}}\qquad
\subfloat[][$q=1$]{\includegraphics[height=5cm]{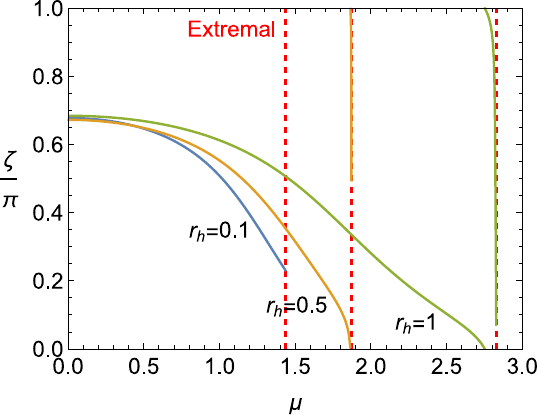}\label{fig:onset_q1_zeta}}\\
\caption{(a) The charged scalar field normal modes of global AdS with constant $A_t=\mu$. (b) The onset of instability of the RNAdS for $r_h=0.1,0.5,1$ at $q=1$.}

\label{fig:onset}
\end{figure}

In figure~\ref{fig:onset}, we show (a) the location of the AdS charged scalar field normal modes~($r_h=0$) and (b) the onset of instability of the RNAdS for $r_h= 0.1, 0.5, 1$ at $q=1$. In figure~\ref{fig:onset}\subref{fig:onset_q1_zeta}, the value of $\mu$ is bounded from above by extremality as $\mu \le \mu_\mathrm{ext}$ \eqref{mu_ext}, which is marked by the vertical red dashed line for each $r_h$. In the same figure, the RNAdS is unstable to the charged scalar field perturbation above each curve, which can be found by studying full quasinormal modes by including nonzero frequencies $\omega$ (see also \cite{Katagiri:2020mvm}). Correspondingly, also in figure~\ref{fig:onset}\subref{fig:onset_rh0}, the scalar field will be nonzero in the region upper from the curve.

In figure~\ref{fig:onset}\subref{fig:onset_rh0}, we emphasize that the normal modes can be characterized by the number of nodes in the radial direction, which increases as the curve reaches $\zeta=0$. The solution without a node is called the fundamental mode, and the solution with nodes are called overtones. Because overtones cost more energy than the fundamental mode, later in the paper, we consider only the backreacted solutions as a fully nonlinear extension of the fundamental mode.

In figure~\ref{fig:onset}\subref{fig:onset_q1_zeta}, the data for $r_h=0.1$ shows that, when the coupling $q$ is small, the onset of instability terminates at the extremality before reaching the Dirichlet boundary conditions ($\zeta=0$). For the Dirichlet boundary conditions to be unstable, a larger $r_h$ is necessary.

\begin{figure}[t]
\centering
\subfloat[][$(r_h,\zeta)$]{\includegraphics[height=5cm]{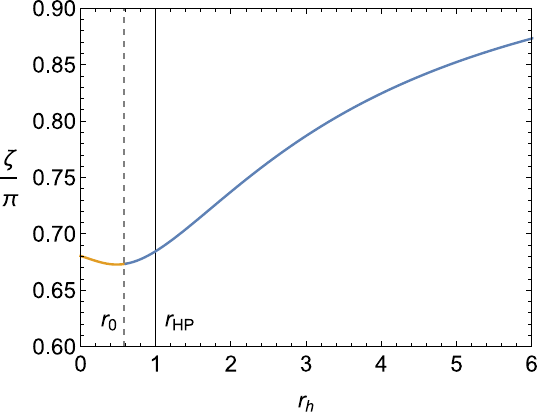}\label{fig:onset_sch_rh}}\qquad
\subfloat[][$(\zeta,T_\mathrm{H})$]{\includegraphics[height=5cm]{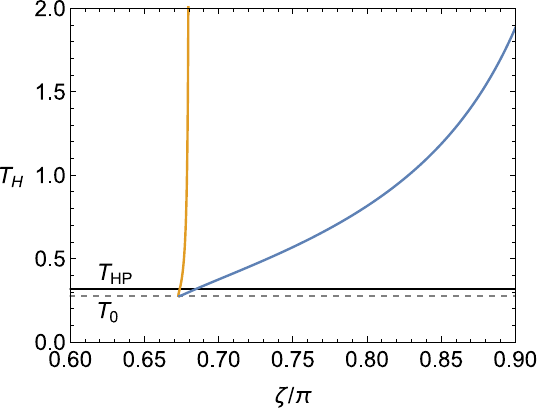}\label{fig:onset_sch_T}}\\
\caption{The onset of instability of the Schwarzschild AdS ($\mu=0$). The blue and orange parts are respectively in the large and small black hole branches of the Schwarzschild AdS. 
Combining with the analysis of quasinormal modes \cite{Katagiri:2020mvm}, we find that, in (a), the Schwarzschild AdS is unstable above the curve, and correspondingly in (b), it is unstable to the right of each of the blue and orange curves.}
\label{fig:onset_sch}
\end{figure}

In figure~\ref{fig:onset_sch}, the onset of instability in the Schwarzschild AdS limit $(\mu=0)$ is shown. The value in the horizonless limit $(r_h=0)$ is analytically given by
\begin{equation}
\zeta_c = \pi-\tan^{-1}(\pi/2) \simeq 0.6805 \pi.
\label{zeta_c}
\end{equation}
In figure~\ref{fig:onset_sch}\subref{fig:onset_sch_rh}, the curve has the minimum at $r_h \simeq 0.4807 (< r_0)$ with $\zeta_\mathrm{min} \simeq 0.6728 \pi$ and approaches $\zeta \to \pi$ as $r_h \to \infty$. There are hence no overtones for the Schwarzschild AdS.

\begin{figure}[t]
\centering
\subfloat[][$(\mu,r_h)$]{\includegraphics[height=5cm]{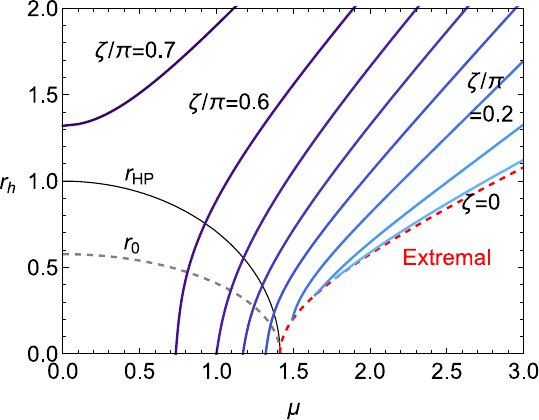}\label{fig:onset_q1_rh}}\qquad
\subfloat[][$(\mu,T_\mathrm{H})$]{\includegraphics[height=5cm]{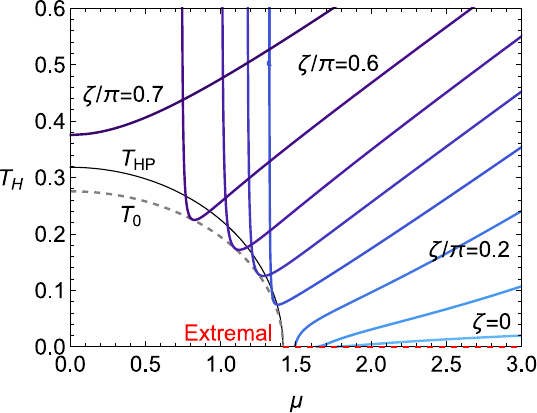}\label{fig:onset_q1_TH}}\\
\caption{The onset of instability of the RNAdS for $q=1$. Color lines show the locations of the onset of instability for $\zeta/\pi=0,0.1,\dots,0.7$ from bottom to top (lighter to darker). For each value of $\zeta$, instability occurs in the region below the corresponding curve.
Panel~(a): phase diagram on the $(\mu, r_H)$ plane. The red dashed line corresponds to the extremal solutions, below which no black hole solutions exist. The curves for $r_h = r_\text{HP}$ (black solid) and $r_h=r_0$ (gray dashed) denote the Hawking-Page transition and the transition between the small/large black holes.
Panel~(b): phase diagram on the $(\mu, T_\mathrm{H})$ plane. The curve for $T_\mathrm{H}=T_0$ (gray dashed) denote the minimum horizon temperature, 
which corresponds to $r_H = r_0$ and no black hole solutions with $T<T_\mathrm{H}$ exist.
Note that $r_h = 0$ on the~$(\mu,r_h)$ plane is mapped to $T_\mathrm{H}\to\infty$ on the~$(\mu, T_\mathrm{H})$ plane.
The red dashed line at $T_\mathrm{H}=0$ corresponds to the extremal solutions.}
\label{fig:onset_q1}
\end{figure}

\begin{figure}[t]
\centering
\subfloat[][$(\mu,r_h)$]{\includegraphics[height=5cm]{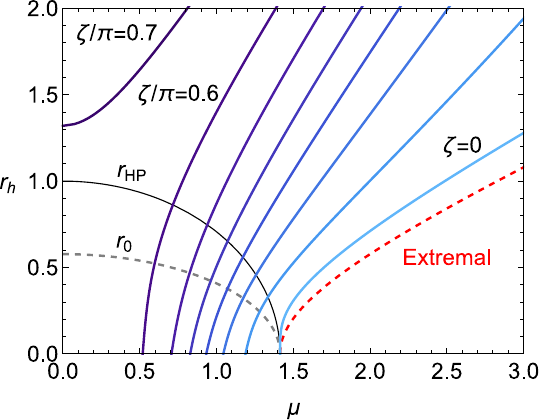}\label{fig:onset_qsqrt2_rh}}\qquad
\subfloat[][$(\mu,T_\mathrm{H})$]{\includegraphics[height=5cm]{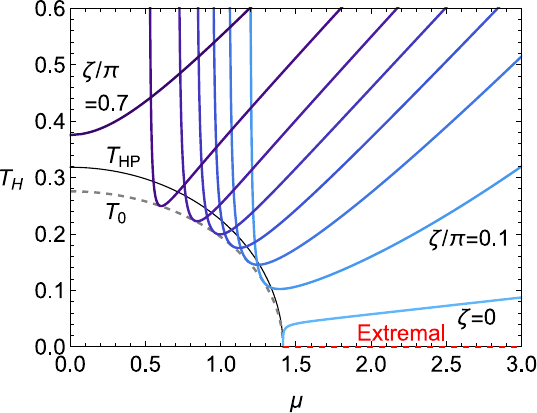}\label{fig:onset_qsqrt2_TH}}\\
\caption{Same as figure~\ref{fig:onset_q1} but for $q=\sqrt{2}$.}
\label{fig:onset_qsqrt2}
\end{figure}

\begin{figure}[t]
\centering
\subfloat[][$(\mu,r_h)$]{\includegraphics[height=5cm]{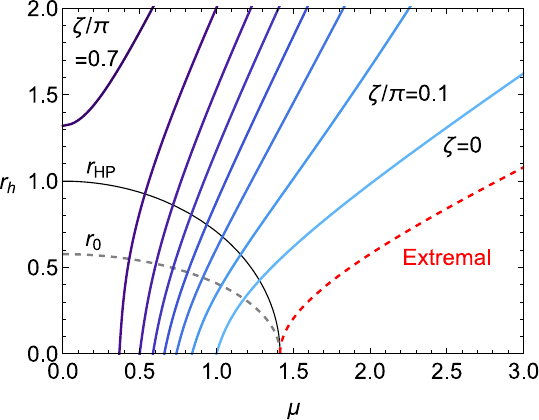}\label{fig:onset_q2_rh}}\qquad
\subfloat[][$(\mu,T_\mathrm{H})$]{\includegraphics[height=5cm]{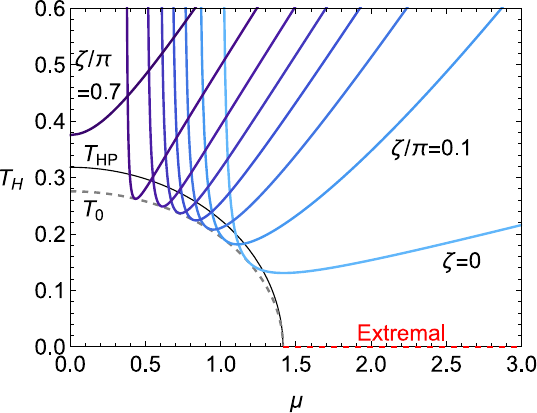}\label{fig:onset_q2_TH}}\\
\caption{Same as figure~\ref{fig:onset_q1} but for $q=2$.}
\label{fig:onset_q2}
\end{figure}

In figures~\ref{fig:onset_q1},\ref{fig:onset_qsqrt2},\ref{fig:onset_q2}, we show the onset of instability of the RNAdS for the fundamental modes with different $\zeta$ at $q=1,\sqrt{2},2$, respectively. The same onset results are shown in the~$(\mu,r_h)$ and $(\mu,T_\mathrm{H})$ planes. We do so because we will discuss the phase structure in the~$(\mu,T_\mathrm{H})$ plane of the phase diagram later in the paper, and it will be instructive to have the location of the instability both in the $(\mu,r_h)$ and $(\mu,T_\mathrm{H})$ planes. 
In each figure, we show the locations of the onset of instability for 8 parameter values $\zeta/\pi=0,0.1,\dots,0.7$. (Among the 8 color lines, the lightest color is $\zeta=0$ and the darkest $\zeta/\pi=0.7$.)
The red dashed line denotes the extremal RNAdS, below which no regular RNAdS exist. The arc by the thin black line ($r_h=r_\mathrm{HP}$ and $T_\mathrm{H}=T_\mathrm{HP}$) is the Hawking-Page transition of the RNAdS \eqref{RN_HP}.
Inside the arc, the thermal AdS is thermodynamically favored over the RNAdS.
In figures~(a), the gray dashed line~($r_h=r_0$) separates the small and large black holes \eqref{RN_minT}, and
small black holes are inside the arc.
In figures (b), the gray dashed line~($T_\mathrm{H}=T_0$) denotes the minimal temperature $T_0$, which is realized when $r_h = r_0$.
In figures~(a), the RNAdS is unstable below each onset curve. In the grand canonical ensemble, we are interested in the onset of instability outside the arc given by $r_h=r_\mathrm{HP}$ or $T_\mathrm{H}=T_\mathrm{HP}$.

Instability can be understood in terms of superradiance~\cite{Katagiri:2020mvm}.\footnote{For charged scalar fields in non-rotating charged black hole spacetimes, superradiance is an amplified scattering of the scalar field around the black hole if its frequency~$\tilde{\omega}$ satisfies $\tilde{\omega} <q\mu$~\cite{PhysRevD.7.949,Hawking:1999dp,Dias:2016pma}. In the context of linear modal stability analysis, the frequency~$\tilde{\omega}$ is replaced by a real part of the quasinormal mode frequency,~${\rm Re}[\tilde{\omega}_{\rm QNM}]$, which is a function of $\zeta$ under the Robin boundary condition. The condition for instability is then ${\rm Re}[\tilde{\omega}_{\rm QNM}]<q\mu$; therefore, instability caused by superradiance is controlled by $\zeta$~(see the details in~\cite{Katagiri:2020mvm}). } With the imposition of the Robin boundary condition, superradiance and the boundary contribution interplay with each other, potentially enhancing instability caused by superradiance depending on the parameters of the scalar field and the background spacetime. It is demonstrated in figures~(a) that, for fixed~$q$ and $r_h$, the value of $\zeta$ at the onset increases as $\mu$ is decreased. That is, the parameter range of $\mu$ for instability is wider as $\zeta$ is increased~(see also figure~8 in~\cite{Katagiri:2020mvm}).

In figure~\ref{fig:onset_q1}, we can see that the extremal RNAdS are stable if $\zeta$ is small and $\mu$ is not sufficiently large, while in figures~\ref{fig:onset_qsqrt2} and \ref{fig:onset_q2}, the extremal RNAdS are unstable to all $\zeta$. 
The critical value for the instability of the Dirichlet boundary condition $\zeta=0$ is $q=\sqrt{2}$, that is, on the phase diagrams on the $(\mu, r_h)$ plane (see Panel (a) of figures~\ref{fig:onset_q1}, \ref{fig:onset_qsqrt2}, \ref{fig:onset_q2}), the onset curve for $\zeta=0$ ends on the (red dashed) curve of the extremal black hole solutions when $q<\sqrt{2}$, while it ends on the $r_h=0$ axis when $q>\sqrt{2}$.\footnote{See \cite{Basu:2016mol} for an earlier discussion on the phase diagram for the Dirichlet boundary condition.
Note that the Hawking-Page transition occurs at $\mu=1$ in this reference due to the normalization different from ours.
See also \cite{Dias:2016pma}, which studied massless scalar but observations given there can be easily generalized to massive cases.}
On the phase diagrams on the $(\mu,T_\mathrm{H})$ plane, the extremal black hole solutions correspond to the $\mu\geq \sqrt{2}$ part of the $T_\mathrm{H}=0$ axis, and only a part of it is covered by the instability region for $\zeta=0$ when $q<\sqrt{2}$, while it is wholly covered by the instability region when $q>\sqrt{2}$.

\subsection{Hairy black holes}

Knowing the onset of instability for the charged scalar field perturbation of the RNAdS, we will construct backreacted hairy black hole solutions branching at the onset of instability. With the ansatz \eqref{metric_ansatz_z}, the equations of motion \eqref{EoMs} are reduced to coupled ODEs for $f(z),\chi(z),\phi(z),A_t(z)$ as
\begin{align}
F'-\left(\frac{3}{z}+z\phi'^2\right)F-e^\chi \left(\frac{z^3 A_t'^2}{2}+\frac{z q^2 A_t^2 \phi^2}{F}\right) + z + \frac{3}{z}+\frac{2 \phi^2}{z}&=0,\label{eom_f}\\
\chi'-2z\phi'^2-\frac{2 z e^\chi q^2 A_t^2 \phi^2}{F^2}&=0,\label{eom_chi}\\
\phi''+\left( \frac{F'}{F}-\frac{\chi'}{2}-\frac{2}{z}\right)+\left(\frac{2}{z^2 F}+\frac{e^\chi q^2 A_t^2}{z^2 F}\right)\phi&=0,\label{eom_phi}\\
A_t''+\frac{\chi'}{2}A_t'-\frac{2 q^2 \phi^2}{z^2 F}A_t&=0,\label{eom_At}
\end{align}
where $F=(1+z^2)f$.

We need the asymptotic behavior of the field variables in $z=0$ and $z = z_h$ or $z \to \infty$. In the AdS boundary $z = 0$, the asymptotic solutions are given by
\begin{align}
f(z) &= 1 + \phi_1^2 z^2 + f_3 z^3 + \left( 2 \phi_1^4 + 2 \phi_2^2 + \frac{a_1^2 e^{\chi_0}}{2} \right) z^4 + \cdots,\label{uv_asymp_f_z}\\
\chi(z) &= \chi_0 + \phi_1^2 z^2 + \frac{8}{3} \phi_1 \phi_2 z^3 + \left( \frac{3}{2} \phi_1^4 + 2 \phi_2^2 - q^2a_0^2 \phi_1^2 e^{\chi_0} \right) z^4 + \cdots,\label{uv_asymp_chi_z}\\
\phi(z) &= \phi_1 z + \phi_2 z^2 + \frac{1}{2} \phi_1 \left( \phi_1^2 - q^2a_0^2 e^{\chi_0} \right) z^3 + \cdots,\label{uv_asymp_phi_z}\\
A_t(z) &= a_0 + a_1 z + q^2 a_0 \phi_1^2 z^2 + \frac{1}{6} \phi_1 \left( 4 q^2 a_0 \phi_2 + (2q^2-1) a_1 \phi_1 \right) z^3 + \cdots,\label{uv_asymp_At_z}
\end{align}
where $(f_3, \chi_0, \phi_1, \phi_2, a_0, a_1)$ are six integration coefficients not determined in the asymptotic analysis. We read off them from the asymptotic form of numerical solutions. With these asymptotic behavior, the metric \eqref{metric_ansatz_z} in $z \to 0$ naively becomes
\begin{equation}
\mathrm{d} s^2|_{z \to 0} = \frac{1}{z^2} \left( -  e^{-\chi_0} \mathrm{d}t^2 + \mathrm{d}z^2 + \mathrm{d}\Omega_2^2 \right).
\end{equation}
This can be rescaled to $\chi_0=0$ by the scaling symmetry (redefinition of $t$) as we will see shortly.

In the presence of the black hole horizon, the regular asymptotic solutions near the horizon $z=z_h=1/r_h$ are given by
\begin{align}
f(z) &= -\frac{3+z_h^2+2\phi_h^2-e^{\chi_h}A_h^2 z_h^3}{z_h(1+z_h^2)}(z-z_h) + \cdots,\label{bh_asymp_f_z}\\
\chi(z) &= \chi_h + \cdots,\label{bh_asymp_chi_z}\\
\phi(z) &= \phi_h + \cdots,\label{bh_asymp_phi_z}\\
A_t(z) &= A_h (z-z_h) + \cdots,\label{bh_asymp_At_z}
\end{align}
where $(\chi_h,\phi_h,A_h)$ are integration constants, and the higher order coefficients are determined fully in terms of them. Two degrees of freedom are considered to be correlated to physical parameters, while the remaining one can be fixed by the scaling symmetry discussed below. In the absence of the horizon, the solutions (\ref{bh_asymp_f_z})--(\ref{bh_asymp_At_z}) are replaced with the following series in $z \to \infty$,
\begin{align}
f(z) &= 1 + O(z^{-2}),\label{ir_asymp_f_z}\\
\chi(z) &= \chi_h + O(z^{-2}),\label{ir_asymp_chi_z}\\
\phi(z) &= \phi_h + O(z^{-2}),\label{ir_asymp_phi_z}\\
A_t(z) &= A_h + O(z^{-2}).\label{ir_asymp_At_z}
\end{align}
There are again three integration constants.

Our ansatz, \eqref{metric_ansatz} and \eqref{At_ansatz}, has the following scaling symmetry:
\begin{equation}
t \to e^{-c/2} t, \quad \chi \to \chi-c, \quad A_t \to e^{c/2} A_t,
\label{scaling_sym}
\end{equation}
where $c$ is an arbitrary constant. By this scaling, solutions with $\chi_0 \neq 0$ can be rescaled to those with canonical boundary metric satisfying $\chi_0 = 0$. This means that, in numerical calculations, we can set the normalization of $\chi$ to an arbitrary value convenient for us without loss of generality. We fix $\chi_h=0$ when we compute and then rescale numerical results by \eqref{scaling_sym} to satisfy $\chi|_{z=0}=0$.

From numerical results, we construct thermodynamic quantities. Carrying out the holographic renormalization as will be described in appendix~\ref{sec:holoreno}, we obtain the expressions of the thermodynamic quantities in terms of the asymptotic coefficients given in (\ref{uv_asymp_f_z})--(\ref{uv_asymp_At_z}). For the Robin boundary conditions, the scalar field is dual to the dimension 1 operator $\mathcal{O}_1$. After rescaling to $\chi_0=0$, the expression of the total energy, charge, and scalar expectation value for the Robin boundary conditions are obtained in \eqref{result_QO1_Robin} and \eqref{result_E_Robin} as (the subscript $R$ is removed here)
\begin{equation}
\begin{split}
\mathcal{E} &= 4 \pi (- f_3 + 3 \phi_1^2 \cot \zeta) = 4 \pi (-f_3 +3 \phi_1 \phi_2), \\
\mathcal{Q} &= - 4 \pi a_1, \qquad \langle \mathcal{O}_1 \rangle = 4 \pi \sqrt{2} \, \phi_1.
\end{split}
\end{equation}
We also have the temperature $T_\mathrm{H}$ through \eqref{T_Hawking} and entropy $\mathcal{S}_\mathrm{BH}\equiv 8 \pi G_N S_\mathrm{BH} = 8 \pi^2 r_h^2$ through \eqref{S_RN}.

We consider the grand canonical ensemble to discuss the phase structure. The grand potential is given by
\begin{equation}
\Omega = \mathcal{E} - T_\mathrm{H} \mathcal{S}_\mathrm{BH} - \mu \mathcal{Q},
\label{quant_stat_rel}
\end{equation}
where $\mu=a_0$. The grand potential $\Omega$ can be evaluated in two different expressions. One is by the combination of thermodynamic quantities as in the RHS of \eqref{quant_stat_rel}, and the other is directly by a bulk integral \eqref{free_energy_robin}. These give the same physical quantity. In practice, the latter is less convenient and costly because of the necessity of numerically cancelling the divergent terms in the integrand. Hence, we use $\Omega $ given by \eqref{quant_stat_rel} when we evaluate the phase structure.

Numerical solutions to (\ref{eom_f})--(\ref{eom_At}) satisfying the Robin boundary conditions can be obtained simply by integrating the equations of motion. Specifying $(\phi_h,A_h,q,r_h)$, we integrate (\ref{eom_f})--(\ref{eom_At}) from the horizon $z=z_h$ (or AdS center $z=\infty)$ to the boundary~$z=0$ and read off $(f_3, \chi_0, \phi_1, \phi_2, a_0, a_1)$ in the asymptotic boundary behavior (\ref{uv_asymp_f_z})--(\ref{uv_asymp_At_z}). After the rescaling to set $\chi_0 \to 0$, we calculate the thermodynamic quantities and $\zeta$ \eqref{Robin_BC}. By these quantities, the grand canonical phase diagram is given as a four-dimensional space $(\mu,T_\mathrm{H},\zeta,q)$. When we present our results, we use data slices in the four dimensional parameter space.

To check numerical results, we can evaluate first-law-like relations generalizing the first law of thermodynamics/black hole mechanics to the case with a nontrivial scalar field. The expressions are discussed in appendix~\ref{sec:1st}. For our solutions in the presence of the scalar field satisfying the Robin boundary conditions, we can use \eqref{1st_R2},
\begin{equation}
\mathrm{d} \mathcal{E} = T_\mathrm{H} \mathrm{d} \mathcal{S}_\mathrm{BH} + \mu \mathrm{d}\mathcal{Q} + \frac{1}{8 \pi} \langle \mathcal{O}_1 \rangle^2 \mathrm{d} ( \cot \zeta).
\label{1st_R2_maintext}
\end{equation}
Note that this contains an atypical variation with respect to $\cot \zeta = \phi_1/\phi_2$, which is not a thermodynamic quantity but is a parameter in the model. However, if we compare between numerical solutions where both $\phi_1$ and $\phi_2$ vary while their ratio is not fixed, the first-law-like equation \eqref{1st_R2_maintext} is useful.
We find that the above relation is satisfied within numerical errors.


\section{Results}
\label{sec:results}

\subsection{Neutral boson stars and black holes}
\label{sec:neutral}

First, we consider neutral solutions.\footnote{See also \cite{Hertog:2004dr,Martinez:2004nb,Hertog:2004ns} for neutral scalar hair solutions with the Robin boundary conditions (in the presence of nonlinear scalar potential).} Here, we focus on the phase transition in the canonical ensemble.\footnote{In a recent work \cite{Basu:2021bpr}, Hawking-Page transition was discussed for neutral black holes with scalar field in the Dirichlet (Neumann) theory when the scalar source was nonzero. Here, this gravitational setup is studied as the double trace deformation with zero scalar source. As we will explain, using the free energy formula for the Robin boundary conditions, we obtain the phase structure comprehensively. Qualitatively similar to \cite{Basu:2021bpr}, we also observe that the Hawking-Page transition temperature increases when the scalar field is nonzero.}

\begin{figure}[t]
\centering
\subfloat[][$\mathcal{E}$]{\includegraphics[height=5cm]{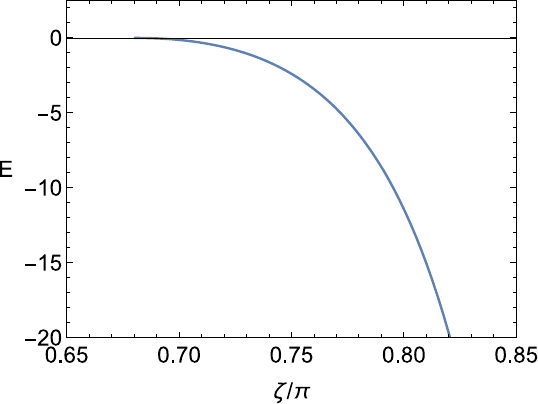}\label{fig:neutral_star_E}}\qquad
\subfloat[][$\langle \mathcal{O}_1 \rangle$]{\includegraphics[height=5cm]{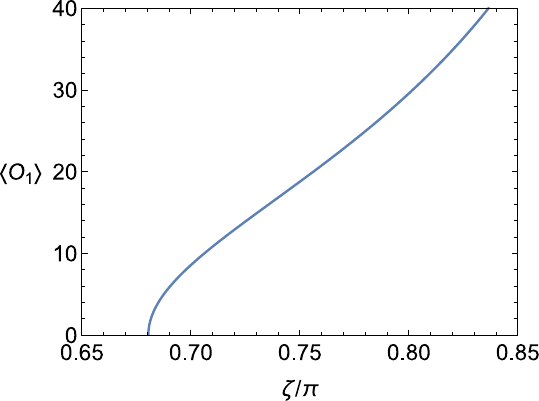}\label{fig:neutral_star_O1}}
\caption{Neutral Robin boson stars.}
\label{fig:neutral_star}
\end{figure}

Before discussing the black holes, let us recall the basic features of the horizonless solutions (see also \cite{Bizon:2020yqs}).
In figure~\ref{fig:neutral_star}, we show the energy and expectation value of neutral horizonless solutions branching at the appearance of the zero normal mode of AdS. Because the scalar field is subject to the Robin boundary conditions, we call the horizonless solutions {\it Robin boson stars}. These are a one-parameter family of solutions parametrized by $\zeta$. Scalar hair grows in $\zeta>\zeta_c\simeq0.6805\pi$, where the phase transition is of second order. The quantities in the figure approach $\langle \mathcal{O}_1 \rangle \to +\infty$ and $\mathcal{E} \to -\infty$ in $\zeta \to \pi$. In the following, we will consider two kinds of generalization: black holes by introducing temperature $T_\mathrm{H}$, and gauge field by adding $(\mu,q)$.

Without the gauge field, the phase structure is specified by two parameters~$(T_\mathrm{H},\zeta)$. In this situation, the free energy we compare for determining the phase structure is nothing but the grand potential \eqref{quant_stat_rel} with $\mu=0$, $\Omega|_{\mu=0}= \left(\mathcal{E} - T_\mathrm{H} \mathcal{S}_\mathrm{BH}\right)_{\mu=0}$. We compare free energies among thermal AdS, Schwarzschild AdS, Robin boson stars, and black holes with neutral scalar hair, which we call {\it Robin black holes}.
The free energy for the thermal AdS is zero, and that for the Schwarzschild AdS is given by \eqref{F_RN} with $\mu=0$.

\begin{figure}[t]
\centering
\subfloat[][Free energy]{\includegraphics[height=5cm]{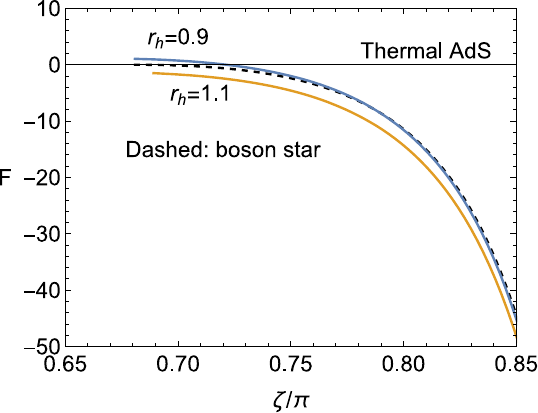}\label{fig:neutral_bh_F}}\qquad 
\subfloat[][Phase diagram]{\includegraphics[height=5cm]{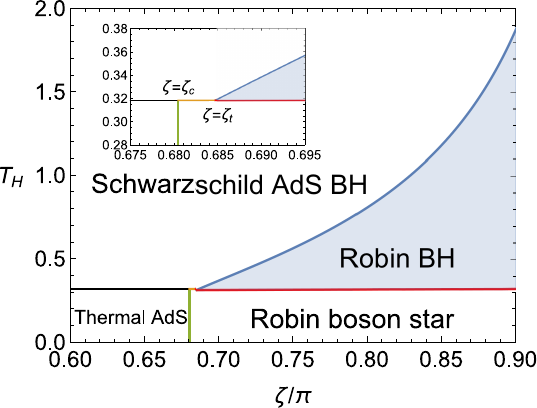}\label{fig:neutral_bh_phase}}
\caption{(a) Comparison of free energies between Robin black holes with $r_h=0.9$ (top, blue) and 1.1 (bottom, orange) and boson stars (black dashed). (b) The phase diagram for neutral solutions.}
\label{fig:neutral}
\end{figure}

In figure~\ref{fig:neutral}\subref{fig:neutral_bh_F}, we show an example of the comparison of free energies among neutral solutions. 
For $r_h \lesssim 1$, the two solutions experience the first order phase transition. In the figure, the lines of the $r_h=0.9$ Robin black holes (blue) and boson stars (black dashed) cross around $\zeta/\pi \sim 0.8$. For $r_h \gtrsim 1$, the free energy of the Robin black holes is always lower than that of the Robin boson stars. The free energy for $r_h=1.1$ (orange) is shown in the figure.

The phase diagram for the neutral solutions is summarized in figure~\ref{fig:neutral}\subref{fig:neutral_bh_phase}. The vertical green line at $\zeta=\zeta_c$ is the second order phase transition from thermal AdS to Robin boson stars. The blue line in $\zeta \ge \zeta_t$ at the border of the Schwarzschild AdS and hairy Robin black holes is the second order phase transition for growing scalar hair, where
\begin{equation}
\zeta_t \simeq 0.6847\pi.
\label{zeta_t}
\end{equation}
Because the source of the scalar field is assumed to be zero, the scalar becomes nonzero spotaneously when the temperature is decreased \cite{Faulkner:2010fh,Faulkner:2010gj}. As $\zeta$ increases, the critical temperature for this scalar hair formation rises, and in the limit $\zeta \to \pi$ ($\cot \zeta \to -\infty$), the Robin black holes dominate at any high temperatures. 
The red line in $\zeta \ge \zeta_t$ marks the first order Hawking-Page transition between Robin black holes and Robin boson stars. The short orange segment in $\zeta_c \le \zeta \le \zeta_t$ (see the inset) is the first order phase transition between Schwarzschild AdS and Robin boson stars; for $\zeta$ in this region, Robin black holes have the higher free energy than these two, and hence the first order phase transition is between the Schwarzschild AdS and Robin boson stars. The three lines (red, orange, blue) merge at $\zeta =\zeta_t$ and
\begin{equation}
T_\mathrm{H} \simeq 0.3184,
\label{T_triplepoint}
\end{equation}
which corresponds to the triple point at which the Schwarzschild AdS black hole, Robin black hole, and the Robin boson star have the same free energy.
The temperature (\ref{T_triplepoint}) at the triple point is slightly higher than the transition temperature $T_\mathrm{HP}$ for the Schwarzschild AdS and thermal AdS phases \eqref{RN_HP}, $T_\mathrm{HP}|_{\mu=0}=1/\pi \simeq 0.3183$.

We find that the Hawking-Page transition temperature depends on $\zeta$ very mildly. We were not able to pin down the line of the Hawking-Page transition up to $\zeta \to \pi$ because of numerical limitations. 
But, as long as we could confirm, the transition temperature (red line) behaves as
\begin{equation}
T_\mathrm{H} \simeq 0.03 (\zeta-\zeta_t)/\pi + 0.3184,
\end{equation}
which is close to $T_\mathrm{HP}|_{\mu=0} \simeq 0.3183$ and
is mostly insensitive to $\zeta$.
Thus, for the Hawking-Page transition temperature of neutral geometries, the effect of the Robin boundary conditions on the free energy is minor.
This behavior suggests that the free energies of the Robin boson star and the Robin black hole changes by almost the same amount when $\zeta$ changes.

\subsection{Charged boson stars}
\label{sec:charged-boson-stars}

To proceed with the reduced number of parameters, we discuss charged but horizonless solutions with the Robin boundary conditions, which we call {\it charged Robin boson stars}. Features of these solutions have been explored in \cite{Gentle:2011kv} in the same setup as ours, but here we discuss the solutions in the phase space parametrized by $(\mu,\zeta,q)$.

\begin{figure}[t]
\centering
\subfloat[][$q=1$]{\includegraphics[height=5cm]{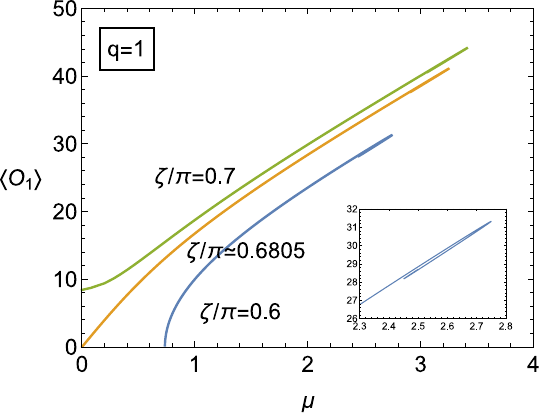}\label{fig:hairy_star_O1_q1}}\qquad 
\subfloat[][$q=2$]{\includegraphics[height=5cm]{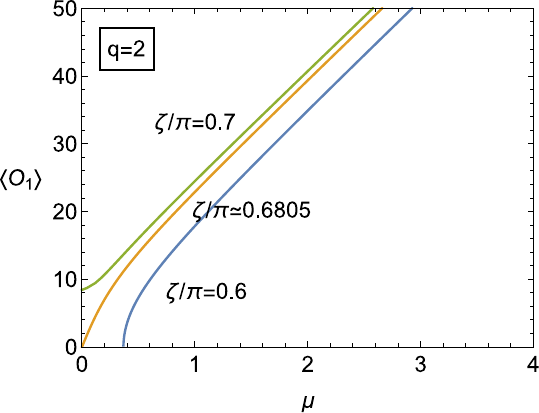}\label{fig:hairy_star_O1_q2}}
\caption{$\langle \mathcal{O}_1 \rangle$ of charged Robin boson stars for $\zeta>\zeta_c$, $\zeta=\zeta_c$, $\zeta<\zeta_c$ at $q=1$ and $2$.}
\label{fig:hairy_star_O1}
\end{figure}

In figure~\ref{fig:hairy_star_O1}, the expectation value $\langle \mathcal{O}_1 \rangle$ is compared for three cases with $\zeta>\zeta_c$, $\zeta=\zeta_c$, and $\zeta<\zeta_c$ for $q=1,2$. Recall that AdS at $\mu=0$ is unstable for $\zeta \ge \zeta_c \simeq 0.6805 \pi$ for forming neutral boson stars. This implies that, for $\zeta > \zeta_c$, charged Robin boson stars are connected to neutral Robin boson stars (with $\mu=0$) by turning on finite~$\mu$.
Meanwhile, for $\zeta < \zeta_c$, they branch at the appearance of the zero normal mode of AdS with finite~$\mu$.
For example, when $\zeta/\pi=0.6$, the value of $\mu$ at the branching point of the condensed solution  in figures~\ref{fig:hairy_star_O1}\subref{fig:hairy_star_O1_q1} and \ref{fig:hairy_star_O1}\subref{fig:hairy_star_O1_q2} (i.e.~the limit of $\langle O_1 \rangle \to 0$) corresponds to $\mu$ in the $r_h=0$ limit in figures~\ref{fig:onset_q1}\subref{fig:onset_q1_rh} and \ref{fig:onset_q2}\subref{fig:onset_q2_rh}, respectively. 
The boundary between these two families of the solutions is $\zeta=\zeta_c$. In addition, in figure~\ref{fig:hairy_star_O1}\subref{fig:hairy_star_O1_q1}, these charged Robin boson stars have the maximal $\mu$ above which solutions do not exist. In the inset, the data region near the maximal $\mu$ for $\zeta/\pi=0.6$ is enlarged.
While it might be visually unclear even in the inset, the region near the largest $\mu$ has a spiral structure, corresponding to the attractor solutions discussed in \cite{Gentle:2011kv}. In figure~\ref{fig:hairy_star_O1}\subref{fig:hairy_star_O1_q2}, the expectation value can be arbitrarily large. This corresponds to solutions allowing the planar limit discussed in \cite{Gentle:2011kv}.

The boundary between these two distinct behaviors depends both on $\zeta$ and $q$. The tendency is that the spiral structure disappears (moves to infinity on the $(\mu, \langle O_1 \rangle)$ plane) as $q$ and $\zeta$ are increased.
Not only $\langle O_1 \rangle$ but also the energy $E$ shows a qualitatively similar behavior.
This tendency can be qualitatively understood as an outcome of the balance between the gravitational attraction, scalar field pressure and the electric repulsion. When $q$ is small, the electric repulsion is weak and then there is a critical mass (and $\langle O_1 \rangle$) for a boson star beyond which the boson star cannot exist. When $q$ is large, the electric repulsion becomes strong enough to sustain the boson star against the gravitational collapse, and correspondingly the mass and $\langle O_1 \rangle$ can become arbitrarily large.

The grand potential of charged Robin boson stars always satisfy $\Omega<0$, where thermal AdS has $\Omega=0$. Therefore, when the charged Robin boson stars exist, they are always preferred over the thermal AdS.
This feature is the same as that in the neutral case, in which the boson stars have the smaller free energy than the thermal AdS (see section~\ref{sec:neutral} and figure~\ref{fig:neutral}\subref{fig:neutral_bh_F}).

\subsection{Charged black holes}

Finally, we consider black holes with nontrivial charged scalar field with the Robin boundary conditions. We call these {\it hairy Robin black holes}. The phase space depends on the all four parameters $(\mu,T_\mathrm{H},\zeta,q)$.

\begin{figure}[t]
\centering
\subfloat[][$\zeta/\pi=0.7$]{\includegraphics[height=5cm]{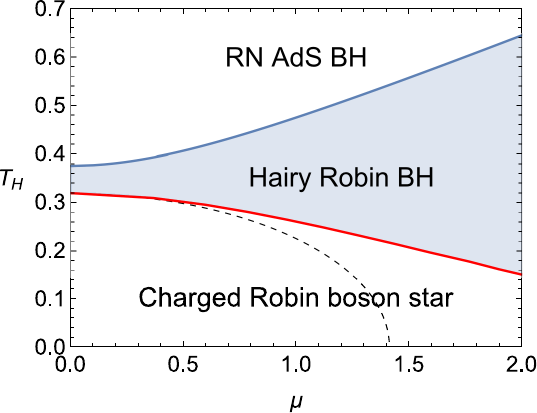}\label{fig:hairy_bh_phase7_q1}}\qquad 
\subfloat[][$\zeta/\pi=0.682$]{\includegraphics[height=5cm]{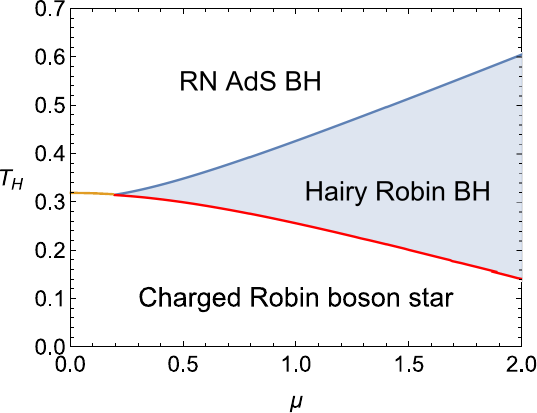}\label{fig:hairy_bh_phase682_q1}}\\
\subfloat[][$\zeta/\pi=0.6$]{\includegraphics[height=5cm]{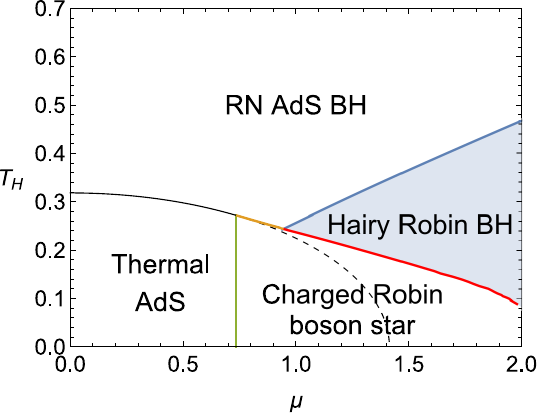}\label{fig:hairy_bh_phase6_q1}}\qquad
\subfloat[][$\zeta/\pi=0.239$]{\includegraphics[height=5cm]{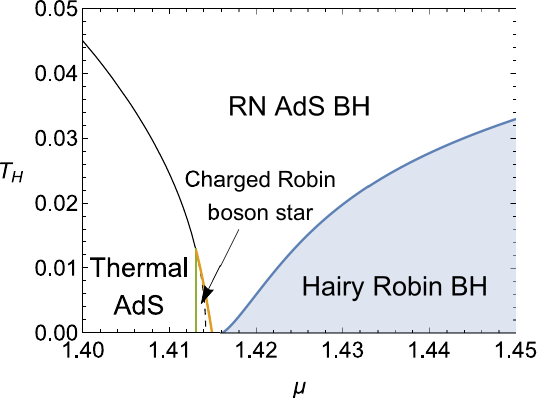}\label{fig:hairy_bh_phase239_q1}}\\
\subfloat[][$\zeta/\pi=0.2$]{\includegraphics[height=5cm]{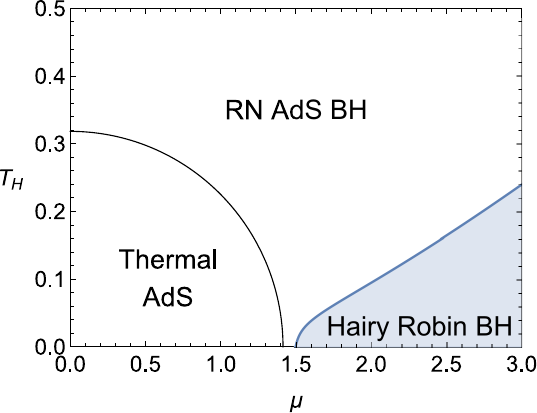}\label{fig:hairy_bh_phase2_q1}}
\caption{Phase diagram for $q=1$ and $\zeta/\pi=0.7, 0.682, 0.6, 0.239, 0.2$.
The dashed and solid black curves denote the Hawking-Page transition temperature $T_\mathrm{HP}$ (Eq.~\ref{RN_minT}) between the RN AdS black holes and the thermal AdS. When the grand potential of these two solutions are bigger than that of the charged Robin boson star, the corresponding part of this curve is irrelevant and does not represent a physical phase boundary, but we added it with a dashed line for reader's convenience.}
\label{fig:hairy_bh_phase_q1}
\end{figure}

In figure~\ref{fig:hairy_bh_phase_q1}, phase diagrams for $q=1$ are shown for different $\zeta$.  In each figure, the blue line on the border between the RNAdS and hairy Robin black holes denotes the second-order phase transition below which the scalar hair forms. The red line is the first order Hawking-Page transition between hairy Robin black holes and charged Robin boson stars. The orange segment denotes the first order phase transition between the RNAdS and charged Robin boson stars. The black dashed line is plotted for reference of the Hawking-Page transition between the thermal AdS and RNAdS \eqref{RN_HP}, although it is not physically dominant because it is superseded by the charged Robin boson star phase.

Starting from a large value of $\zeta$, we browse notable features in the phase structure by decreasing $\zeta$.
\begin{itemize}
\item $\zeta>\zeta_t\simeq 0.6847\pi$: In figure~\ref{fig:hairy_bh_phase_q1}\subref{fig:hairy_bh_phase7_q1} (see Eq.~(\ref{zeta_t}) for the definition of $\zeta_t$), neutral solutions ($\mu=0)$ can have nontrivial scalar hair. Thermal AdS does not appear because its free energy is always higher than Robin boson stars when the latter exist as solutions. Hence, the phase diagram contains three phases: zero scalar RNAdS, hairy Robin black holes, and charged Robin boson stars. By decreasing the temperature, the RNAdS spontaneously grows the scalar hair, and then the hairy Robin black hole transitions to the charged Robin boson stars. This feature is common to all $\mu$.
\item $\zeta_t > \zeta > \zeta_c\simeq 0.6805\pi $: In figure~\ref{fig:hairy_bh_phase_q1}\subref{fig:hairy_bh_phase682_q1}, the phase structure for this parameter region is shown for $\zeta/\pi=0.682$. When $\zeta$ is decreased to $\zeta_t$, the two phase transition lines~(blue and red) first meet at $\mu=0$.
As shown in figure~\ref{fig:neutral}\subref{fig:neutral_bh_phase}, 
$\zeta=\zeta_t$ is bigger than $\zeta=\zeta_c$ where the thermal AdS phase shows up. This means that, in $\zeta < \zeta_t$, the phase transition from the RNAdS to charged Robin boson stars (orange line) appears. 
\item $\zeta_c> \zeta \gtrsim 0.24\pi $: In $\zeta<\zeta_c$, the thermal AdS phase can be present as $\mu$ is increased from $0$ until the charged Robin boson stars branch from thermal AdS as discussed in figure~\ref{fig:hairy_star_O1}. The phase diagram in this parameter region is shown in figure~\ref{fig:hairy_bh_phase_q1}\subref{fig:hairy_bh_phase6_q1}. 
The vertical green line is the second order phase transition between thermal AdS and charged Robin boson stars.
\item In figures~\ref{fig:hairy_bh_phase_q1}\subref{fig:hairy_bh_phase7_q1}--\ref{fig:hairy_bh_phase_q1}\subref{fig:hairy_bh_phase6_q1}, the Hawking-Page transition (red line) will approach $T_\mathrm{H}\to 0$ as $\mu$ is increased. We were not able to compute until this limit due to tough numerics, but we can see that the transition line will go down towards $T_\mathrm{H} \to 0$ for a wide parameter range (in $\zeta/\pi \gtrsim 0.24$). 
We also expect that the Hawking-Page transition should go to $T_\mathrm{HP} \to 0$ before boson star solutions disappear
at the upper limit in $\mu$ for Robin boson stars with small $q$ (discussed in section~\ref{sec:charged-boson-stars}).
\item $\zeta \simeq 0.24 \pi$: When $\zeta$ is decreased further, the Hawking-Page transition between hairy Robin black holes and charged Robin boson stars reaches zero temperature and disappears. For $q=1$, this occurs in a small parameter window of $\zeta$ near $\zeta/\pi \simeq 0.24$.
Figure~\ref{fig:hairy_bh_phase_q1}\subref{fig:hairy_bh_phase239_q1} is the phase diagram for $\zeta/\pi=0.239$. This has four phases, but the charged Robin boson stars and hairy Robin black holes are separated by the RNAdS, and correspondingly there is a small gap of $\mu$ where the extremal RNAdS survives in the phase diagram at zero temperature. The hairy Robin black holes branch from the extremal RNAdS.
\item $\zeta \lesssim 0.24 \pi$: The charged Robin boson star phase then disappears when $\zeta$ is decreased further. In figure~\ref{fig:hairy_bh_phase_q1}\subref{fig:hairy_bh_phase2_q1}, the phase diagram at $\zeta/\pi=0.2$ is shown. While charged Robin boson stars also exist as solutions in this parameter region, their grand potential is always bigger than that of hairy Robin black holes, and hence they do not show up in the grand canonical phase diagram.
\end{itemize}

When the coupling $q$ is increased, the $\zeta$ dependence of the phase structure can be different.
\begin{itemize}
\item For $q=\sqrt{2}$, the phase structures of figures~\ref{fig:hairy_bh_phase_q1}\subref{fig:hairy_bh_phase7_q1}, \ref{fig:hairy_bh_phase_q1}\subref{fig:hairy_bh_phase682_q1}, and~\ref{fig:hairy_bh_phase_q1}\subref{fig:hairy_bh_phase6_q1} are observed for $\zeta>0$, but those of figures~\ref{fig:hairy_bh_phase_q1}\subref{fig:hairy_bh_phase239_q1} and~\ref{fig:hairy_bh_phase_q1}\subref{fig:hairy_bh_phase2_q1} are absent because no stable extremal RNAdS exist even for the Dirichlet boundary condition~$\zeta=0$. Instead, at $\zeta=0$, a phase structure not shown here appears (see figure~7(a) in~\cite{Basu:2016mol}). It contains three phases, where thermal AdS and hairy Robin BH are separated by the RN AdS BH reaching sufficiently low temperature. There, the phase of the charged boson stars also disappears because the onset is exactly on the $T_\mathrm{H}=0$ axis~\cite{Basu:2016mol}.

\item For $q > \sqrt{2}$, the scalar hair grows at finite temperatures before extremality is reached, because all the extremal solutions with $T_\mathrm{H}=0$ are unstable toward scalar hair formation when $q>\sqrt{2}$, as explained in section~\ref{sec:instability-of-RNAdS}.
Therefore, the phase structures depicted in figures~\ref{fig:hairy_bh_phase_q1}\subref{fig:hairy_bh_phase239_q1} and \ref{fig:hairy_bh_phase_q1}\subref{fig:hairy_bh_phase2_q1} are absent in $q > \sqrt{2}$.

\end{itemize}

There is qualitative difference for the phase structures of the Robin boundary conditions from those for the Dirichlet boundary condition (see \cite{Basu:2016mol}) that some of the phase structures in figure~\ref{fig:hairy_bh_phase_q1} are absent in the same system under the Dirichlet boundary condition.
The structures of figures~\ref{fig:hairy_bh_phase_q1}\subref{fig:hairy_bh_phase7_q1} and \ref{fig:hairy_bh_phase_q1}\subref{fig:hairy_bh_phase682_q1} 
do not exist
for the Dirichlet boundary condition, 
because thermal AdS phase
should appear 
in small $\mu$ region 
when $\zeta < \zeta_c$
and particularly in the Dirichlet case ($\zeta=0$).
Adding to that, the presence of neutral Robin black hole phase (with $\mu=0$) for $\zeta>\zeta_t$ observed in figure~\ref{fig:hairy_bh_phase_q1}\subref{fig:hairy_bh_phase7_q1} is another unique feature of the Robin boundary condition.
The structure of figure~\ref{fig:hairy_bh_phase_q1}\subref{fig:hairy_bh_phase6_q1} is observed for the Dirichlet boundary condition with a gauge coupling $q>\sqrt{2}$ (in our normalization)~\cite{Basu:2016mol}. For the Robin boundary condition, however, this phase structure can be seen even for small $q$ if $\zeta$ is sufficiently large. The structure of figure~\ref{fig:hairy_bh_phase_q1}\subref{fig:hairy_bh_phase239_q1} is not seen for the Dirichlet boundary condition because the charged boson star phase disappears at the same time as $T_\mathrm{HP} \to 0$ at $q=\sqrt{2}$ (see~\cite{Basu:2016mol}). The structure of figure~\ref{fig:hairy_bh_phase_q1}\subref{fig:hairy_bh_phase2_q1} is typical in $q<\sqrt{2}$.

\section{Conclusion \label{sec:conclusion}}

We considered charged boson stars and black holes in four-dimensional Einstein-Maxwell-complex scalar theory with the Robin boundary conditions for the charged scalar field in asymptotically global AdS spacetime. This setup is dual to the double trace deformation of three-dimensional dual field theory on $R \times S^2$ with a dimension 1 charged scalar operator. The current setup has the four-dimensional parameter space $(T_\mathrm{H},\mu,q,\zeta)$, and the consideration of the Robin boundary conditions offers the most general solutions in the four-dimensional Einstein-Maxwell-complex scalar theory. The phase structure and phase transition are studied in the grand canonical ensemble. There are four phases characterized by the presence and absence of the black hole horizon and nontrivial scalar hair. 
There is an interplay between two kinds of instability on the formation of a charged scalar hair, the one caused by the Robin boundary conditions and the other by the chemical potential or the black hole charge.
These introduce the richer phase structure
compared with the case of the Dirichlet boundary condition,
as explained in section~\ref{sec:results}.

We considered the Robin boundary conditions for scalar field in this paper. This type of boundary conditions can be also imposed on vector and metric fields \cite{Ishibashi:2004wx,Witten:2003ya,Marolf:2006nd,Compere:2008us}. It will be interesting to consider phases of gravitational solutions where the Robin boundary conditions are imposed on such different kinds of field.
Rather recently, the Robin boundary conditions are utilized for studies in various context including the holography and also the supergravity (see e.g.\ \cite{Ahn:2022azl,Jeong:2023las,Nozawa:2022upa}).
Our study would provide useful information to clarify various properties in these cases, such as the thermodynamical phase structures and also the dynamical (in)stabilities.

\acknowledgments

The authors thank Li Li, Keiju Murata and Matthew Roberts for useful discussions,
and also thank the JGRG webinar series where this work was initiated.
The work of T.H. was supported in part by 
JSPS KAKENHI Grant Numbers 19H01895, 20H05853, and 19K03876.
The work of T.I.\ was supported in part by JSPS KAKENHI Grant Number 19K03871.
The work of T.K. was supported in part by 
JSPS KAKENHI Grant Number 17H06360. T.K. thanks for support by VILLUM FONDEN (grant no. 37766), by the Danish Research Foundation, and under the European Union's H2020 ERC Advanced Grant ``Black holes: gravitational engines of discovery'' grant agreement no. Gravitas–101052587.
The work of N.T.\ was supported in part by JSPS KAKENHI Grant Numbers 18K03623 and 21H05189.

\appendix

\section{Holographic renormalization}
\label{sec:holoreno}

We carry out holographic renormalization in the asymptotically global AdS spacetime with the Robin boundary conditions (also called the mixed boundary conditions) \cite{Papadimitriou:2007sj}. We follow the calculations in \cite{Caldarelli:2016nni}, the application of which to complex scalar theory in global AdS is straightforward.

We use the $r$-coordinate in calculation. The asymptotic solutions near the AdS boundary (\ref{uv_asymp_f_z})--(\ref{uv_asymp_At_z}) take the form
\begin{align}
f(r) &= 1 + \frac{\phi_1^2}{r^2} + \frac{f_3}{r^3} + \left( 2 \phi_1^4 + 2 \phi_2^2 + \frac{a_1^2 e^{\chi_0}}{2} \right) \frac{1}{r^4} + \cdots,\label{uv_asymp_f_r}\\
\chi(r) &= \chi_0 + \frac{\phi_1^2}{r^2} + \frac{8}{3} \frac{\phi_1 \phi_2}{r^3} + \left( \frac{3}{2} \phi_1^4 + 2 \phi_2^2 - q^2 a_0^2 \phi_1^2 e^{\chi_0} \right) \frac{1}{r^4} + \cdots,\label{uv_asymp_chi_r}\\
\phi(r) &= \frac{\phi_1}{r} + \frac{\phi_2}{r^2} + \frac{\phi_1}{2} \left( \phi_1^2 - q^2 a_0^2 e^{\chi_0} \right) \frac{1}{r^3} + \cdots,\label{uv_asymp_phi_r}\\
A_t(r) &= a_0 + \frac{a_1}{r} + \frac{q^2 a_0 \phi_1^2}{r^2} + \frac{1}{6} \phi_1 \left( 4 q^2 a_0 \phi_2 + (2q^2-1) a_1 \phi_1 \right) \frac{1}{r^3} + \cdots.\label{uv_asymp_At_r}
\end{align}
In the following, we assume that the scaling \eqref{scaling_sym} has been applied so that $\chi_0=0$.

The action is regularized by introducing a cutoff surface at $r=r_\Lambda$. Let $M$ denote the regularized spacetime manifold defined in $r \le r_\Lambda$ and $\partial M$ the cutoff surface at $r=r_\Lambda$. The bulk action \eqref{EMS_action} accompanied by the Gibbons-Hawking term can be regularized as
\begin{equation}
S_\mathrm{reg} = \frac{1}{8 \pi G_N} \int_M \mathrm{d}^4 x \sqrt{-g} \left(\frac{1}{2}\left( R - 2 \Lambda \right) - \frac{1}{4} F_{\mu\nu} F^{\mu\nu} - |D \phi|^2 - m^2 |\phi|^2 \right)
+ S_\mathrm{GH},
\label{S_reg}
\end{equation}
where
\begin{equation}
S_\mathrm{GH} = \frac{1}{8 \pi G_N} \int_{\partial M} \mathrm{d}^3 x \sqrt{-\gamma} \, K
\label{S_GH}
\end{equation}
with $K \equiv K_{ij} \gamma^{ij}$ being the trace of the extrinsic curvature $K_{ij}$ with respect to the induced metric $\gamma_{ij}$ on $\partial M$ ($i,j$ run over the three-dimensional coordinates on $\partial M$). The extrinsic curvature is given by
\begin{equation}
K_{ij} = \frac{1}{2} \delta_i^\mu \delta_j^\nu \left( \nabla_\mu n_\nu + \nabla_\nu n_\mu \right),
\end{equation}
where $n^\mu$ is an outward unit normal $g_{\mu\nu} n^\mu n^\nu = 1$. However, the ``bare'' action \eqref{S_reg} diverges when the cutoff is simply removed by taking the limit of $r_\Lambda \to \infty$.

This divergence can be cancelled by counterterms $S_\mathrm{ct}$. Including $S_\mathrm{ct}$ formally, we can define a subtracted action that is finite in the limit $r_\Lambda \to \infty$ as
\begin{equation}
S_\mathrm{sub}=S_\mathrm{reg} + S_\mathrm{ct}.
\label{S_sub}
\end{equation}
Then, removing the cutoff gives a renormalized action,
\begin{equation}
S_\mathrm{ren} = \lim_{r_\Lambda \to \infty} S_\mathrm{sub}.
\label{S_ren}
\end{equation}
The form of $S_\mathrm{ct}$ depends on the boundary conditions at the AdS boundary. We will discuss the cases of the Dirichlet theory, Neumann theory, and double trace deformation in turn.

\paragraph{Dirichlet theory}
When $\zeta=0$, our Einstein-Maxwell complex scalar system is treated as the Dirichlet theory that has a dimension 2 operator $O_2$ in the dual field theory on the AdS boundary. This is also called the standard quantization. The counterterms for this case can be given by \cite{Balasubramanian:1999re,deHaro:2000vlm,Caldarelli:2016nni}
\begin{equation}
S_\mathrm{ct} = - \frac{1}{8 \pi G_N}\int_{\partial M} \mathrm{d}^3x \sqrt{-\gamma} \left[ 2 + \frac{R_\gamma}{2} + \phi^2 \right],
\label{S_ct}
\end{equation}
where $R_\gamma$ is the Ricci scalar for $\gamma_{ij}$, and we ignored derivative terms of the scalar field that do not contribute in our spherically symmetric static solutions. With these counterterms, let $S_\mathrm{ren}^D$ denote the renormalized action for the Dirichlet theory.

The expectation values of field theory operators can be obtained through variation as
\begin{equation}
\delta S_\mathrm{ren}^D = \int \mathrm{d}^3x \sqrt{-h} \left( \frac{1}{2} \langle T^{ij} \rangle \delta h_{ij} + \langle J^i \rangle \delta \Psi_i + \langle O_2 \rangle \delta \Phi_D \right),
\label{S_ren_var_D}
\end{equation}
where $\Phi_D = \sqrt{2} \phi_1$ is the source of the scalar operator, $\Psi_i$ denotes that of the gauge field, and $h_{ij}$ are the metric components of the boundary $R \times S^2$. For the gauge field, we turn on the chemical potential $\Psi_t = \mu = a_0$.

The boundary stress energy tensor can be practically calculated as follows. From the subtracted action, the stress energy tensor on the cutoff surface can be obtained as
\begin{equation}
(T_\gamma)_{ij} = - \frac{2}{\sqrt{-\gamma}} \frac{\delta S_\mathrm{sub}^D}{\delta \gamma^{ij}} = \frac{1}{8 \pi G_N} \left(- K_{ij} + K \gamma_{ij} - 2 \gamma_{ij} + (G_\gamma)_{ij} - \phi^2 \gamma_{ij} \right),
\label{T_gamma_ij}
\end{equation}
where $(G_\gamma)_{ij} = (R_\gamma)_{ij} - \frac{1}{2} R_\gamma \gamma_{ij}$ is the Einstein tensor for the induced metric. This scales as $(T_\gamma)_{ij} \sim 1/r_\Lambda$ because $\gamma^{ij} \sim 1/r_\Lambda^2$ and $\sqrt{-\gamma} \sim r_\Lambda^3$. Hence, by switching from $\gamma_{ij}$ to $h_{ij}$, the expectation value of the boundary stress energy tensor \eqref{S_ren_var_D} reads
\begin{equation}
\langle T_{ij} \rangle = \lim_{r_\Lambda \to \infty} r_\Lambda (T_\gamma)_{ij}.
\end{equation}
Explicitly, the components are given by
\begin{align}
8 \pi G_N \langle T_{tt} \rangle &= - f_3 + 2 \phi_1 \phi_2,  \label{Ttt_Dirichlet}\\
8 \pi G_N \langle T_{\theta\theta} \rangle &= - \frac{f_3}{2} + 2 \phi_1 \phi_2,  \label{Tthth_Dirichlet}\\
8 \pi G_N \langle T_{\psi\psi} \rangle &= \sin^2 \theta \left(- \frac{f_3}{2} + 2 \phi_1 \phi_2 \right), \label{Tpsps_Dirichlet}
\end{align}
where $(\theta,\psi)$ denote the coordinates on $S^2$ introduced as $\mathrm{d}\Omega_2^{2} = \mathrm{d}\theta^2 + \sin^2 \! \theta \mathrm{d}\psi^2$. From the stress energy tensor, the total energy (also called the total mass) of the Dirichlet theory is expressed as
\begin{equation}
\mathcal{E}_D \equiv 8 \pi G_N \int \mathrm{d}\Omega_2 \langle T_{tt} \rangle = 4 \pi (-f_3 +2 \phi_1 \phi_2),
\label{Dirichlet_E}
\end{equation}
where $8 \pi G_N$ is included in the definition of the LHS.

The expectation values for the matter fields are
\begin{equation}
8 \pi G_N \langle J^t \rangle = - a_1, \quad
8 \pi G_N \langle O_2 \rangle = \sqrt{2} \phi_2.
\end{equation}
These quantities are the densities per solid angle. The total charge is given by
\begin{equation}
\mathcal{Q}\equiv 4 \pi \cdot 8 \pi G_N \langle J^t \rangle = - 4 \pi a_1.
\end{equation}
Similarly, the scalar expectation value integrated over the sphere is
\begin{equation}
\langle \mathcal{O}_2 \rangle \equiv 4 \pi \cdot 8 \pi G_N \langle O_2 \rangle = 4 \pi \sqrt{2} \, \phi_2 .
\label{Dirichlet_O2}
\end{equation}

The trace of the stress energy tensor satisfies
\begin{equation}
\langle T^i{}_i \rangle = \frac{2 \phi_1 \phi_2}{8 \pi G_N}  = \Phi_D \langle O_2 \rangle.
\label{Dirichlet_Tii_trace}
\end{equation}
If both $\phi_1$ and $\phi_2$ are nonzero, the theory that gives the variation \eqref{S_ren_var_D} can be interpreted as Dirichlet theory in the presence of a nonzero source $\Phi_D$. The nonzero trace \eqref{Dirichlet_Tii_trace} then indicates that the conformal symmetry is explicitly broken by the source. When the source is absent $\Phi_D=0$, i.e.~$\zeta=0$, the expression of the energy \eqref{Dirichlet_E} reduces to $\mathcal{E}_D|_{\phi_1=0}=-4 \pi f_3$.

We can also calculate the finite Euclidean on-shell action when the counterterms are added.\footnote{We thank Li Li for discussions on this calculation.} Using the equations of motion, we obtain (see~section 3.4 in \cite{Hartnoll:2008kx})
\begin{equation}
\frac{1}{2} \left(R-2\Lambda\right) + \mathcal{L} = -\frac{1}{2} \left(G^t{}_t+G^r{}_r\right) = -\frac{1}{\sqrt{-g}}\left( \frac{(1+r^2) f \sqrt{-g}}{r} \right)^\prime + \frac{1}{r^2},
\end{equation}
where the last term, which is not a total derivative, is due to the spherical topology of the global AdS. By this relation, the bulk action \eqref{EMS_action} is simplified to
\begin{align}
S_\mathrm{bulk} &= \frac{4 \pi}{8 \pi G_N} \int \mathrm{d}t \mathrm{d}r \left[ \left( -r (1+r^2) f e^{-\frac{\chi}{2}} \right)^\prime + e^{-\frac{\chi}{2}} \right] \nonumber \\
&=\frac{4 \pi}{8 \pi G_N} \int \mathrm{d}t \left[ -r_\Lambda^3 - \left(1+\frac{\phi_1^2}{2}\right)r_\Lambda - f_3 + \frac{4 \phi_1 \phi_2}{3} + O\left(\frac{1}{r_\Lambda} \right) + \int^{r_\Lambda}_{r_h} \mathrm{d}r \, e^{-\frac{\chi}{2}} \right],
\label{S_bulk_onshell}
\end{align}
where \eqref{uv_asymp_f_r}--\eqref{uv_asymp_At_r} were used.
This diverges for $r_\Lambda\to\infty$.
The divergence can be cancelled by adding the counterterms as well as the Gibbons-Hawking term,
\begin{equation}
S_\mathrm{GH}+S_\mathrm{ct} = \frac{4 \pi}{8 \pi G_N} \int \mathrm{d}t \left[ r_\Lambda^3 + \frac{\phi_1^2}{2} r_\Lambda + \frac{f_3}{2} + \frac{2 \phi_1 \phi_2}{3} + O\left(\frac{1}{r_\Lambda} \right) \right].
\label{S_GH_ct_onshell}
\end{equation}
Combining \eqref{S_bulk_onshell} and \eqref{S_GH_ct_onshell}, we obtain the finite Lorentzian renormalized on-shell action,
\begin{equation}
S_L = S_\mathrm{ren}^D = \lim_{r_\Lambda \to \infty} \left( S_\mathrm{bulk}+S_\mathrm{GH}+S_\mathrm{ct} \right).
\end{equation}
The Euclidean on-shell action $S_E$ can be obtained by replacing $\int \mathrm{d}t \to - \int_0^{1/T_\mathrm{H}} d \tau$ where $\tau$ denotes the Euclidean time. It is related to the grand potential for the Dirichlet theory as $\Omega_D \equiv 8 \pi G_N T_\mathrm{H} S_E$. The expression of the grand potential in terms of the bulk integral is hence given by
\begin{equation}
\Omega_D = 4 \pi \int_{r_h}^\infty \mathrm{d}r (1-e^{-\frac{\chi}{2}}) + 4 \pi \left( \frac{f_3}{2} - 2 \phi_1 \phi_2 + r_h \right),
\label{free_energy_dirichlet}
\end{equation}
where we used $r_\Lambda = \int_{r_h}^{r_\Lambda} \mathrm{d}r + r_h$ to rewrite the cutoff dependence for numerical evaluation of the $r$-integral.

\paragraph{Neumann theory}
For $\zeta \neq 0$, the bulk theory is considered to be dual to the boundary field theory with a dimension 1 scalar operator $O_1$. This is known as the alternative quantization. The case of $\zeta=\pi/2$ is the Neumann theory. It turns out that the source of the scalar operator is identified as $\Phi_N = -\sqrt{2} \phi_2$, and the expectation value of the scalar operator is $8 \pi G_N \langle O_1 \rangle = \sqrt{2} \phi_1$. The renormalized action is modified from the Dirichlet theory as follows.

The Neumann theory is the Legendre transform of the Dirichlet theory \cite{Klebanov:1999tb},
\begin{equation}
S_\mathrm{ren}^N = S_\mathrm{ren}^D + S_\mathrm{LT},
\label{S_Neumann}
\end{equation}
where $N$ denotes the Neumann theory, and
\begin{equation}
S_\mathrm{LT} = - \frac{2}{8 \pi G_N} \int \mathrm{d}^3x \sqrt{-h} \, \phi_1 \phi_2 .
\label{S_DN_Legendre}
\end{equation}
The variation with respect to the scalar field gives
\begin{equation}
\delta _\phi S_\mathrm{LT} = - \frac{2}{{8 \pi G_N}} \int \mathrm{d}^3x \sqrt{-h} \left( \phi_2 \delta \phi_1 + \phi_1 \delta \phi_2 \right) = \int \mathrm{d}^3x \sqrt{-h} \left( -\langle O_2 \rangle \delta \Phi_D + \langle O_1 \rangle \delta \Phi_N \right).
\end{equation}
The variation of the renormalized Neumann action hence takes the form
\begin{equation}
\delta S_\mathrm{ren}^N = \int \mathrm{d}^3x \sqrt{-h} \left( \frac{1}{2} \langle T^{ij} \rangle \delta h_{ij} + \langle J^i \rangle \delta \Psi_i + \langle O_1 \rangle \delta \Phi_N \right).
\label{S_ren_var_N}
\end{equation}
In the above equation, $\langle T^{ij} \rangle \delta h_{ij}$ contains the contribution from the variation of \eqref{S_DN_Legendre} by $h_{ij}$, which shifts (\ref{Ttt_Dirichlet})--(\ref{Tpsps_Dirichlet}). The stress energy tensor for the Neumann theory is thus given by
\begin{align}
8 \pi G_N \langle T_{tt} \rangle &= - f_3 + 4 \phi_1 \phi_2, \label{Ttt_Neumann}\\
8 \pi G_N \langle T_{\theta\theta} \rangle &= - \frac{f_3}{2}, \label{Tthth_Neumann}\\
8 \pi G_N \langle T_{\psi\psi} \rangle &= - \sin^2 \theta \, \frac{f_3}{2}. \label{Tpsps_Neumann}
\end{align}
The trace of the stress energy tensor is
\begin{equation}
\langle T^i{}_i \rangle = -\frac{4 \phi_1 \phi_2}{8 \pi G_N}  = 2 \Phi_N \langle O_1 \rangle.
\label{Neumann_Tii_trace}
\end{equation}
Correspondingly, the total energy is
\begin{equation}
\mathcal{E}_N = 4 \pi (-f_3 + 4 \phi_1 \phi_2) = \mathcal{E}_D + \mathcal{E}_\mathrm{LT},
\label{Neumann_E}
\end{equation}
and so is the grand potential, $\Omega_N = \Omega_D + \Omega_\mathrm{LT}$, where $  \mathcal{E}_\mathrm{LT} = \Omega_\mathrm{LT} = 8 \pi \phi_1 \phi_2$. When the source is absent $\Phi_N=0$, i.e.~$\zeta=\pi/2$, the energy is given by $\mathcal{E}_N|_{\phi_2=0}=-4 f_3$.

\paragraph{Double trace deformation} For $\zeta \neq 0$ nor $\pi/2$, the theory is interpreted as double trace deformation of the Neumann theory. For this, we need to include additional finite boundary terms in order for consistent variation with respect to the source in the deformed theory. We give the source in the form
\begin{equation}
\Phi_R = - \sqrt{2} \left( \phi_2 + \alpha \phi_1 \right),
\label{alphadef}
\end{equation}
where $\alpha$ is a real parameter. The undeformed Neumann theory corresponds to $\alpha=0$.
For this source, we need an additional finite boundary term,
\begin{equation}
S_\mathrm{Dtr} = -\frac{\alpha}{8 \pi G_N} \int \mathrm{d}^3x \sqrt{-h} \, \phi_1^2.
\label{S_F_finite}
\end{equation}
This term corresponds to the relevant double trace deformation of the dual field theory. The renormalized action is modified to 
\begin{equation}
S_\mathrm{ren}^R = S_\mathrm{ren}^N + S_\mathrm{Dtr} = S_\mathrm{ren}^D + S_\mathrm{LT} + S_\mathrm{Dtr}.
\label{S_ren_double}
\end{equation}

The renormalized action equipped with the finite term $S_\mathrm{Dtr}$ gives the correct variation with respect to the source $\Phi_R$. The scalar field variation of \eqref{S_ren_double} is
\begin{equation}
\delta_\phi S_\mathrm{ren}^R = \frac{1}{8 \pi G_N} \int \mathrm{d}^3x \sqrt{-h} \left( -2 \phi_1 \delta \phi_2 - 2 \alpha \phi_1 \delta \phi_1 \right) = \int \mathrm{d}^3x \sqrt{-h} \langle O_1 \rangle \delta \Phi_R.
\end{equation}
The full variation of $S_\mathrm{ren}^R$ takes the form
\begin{equation}
\delta S_\mathrm{ren}^R = \int \mathrm{d}^3x \sqrt{-h} \left( \frac{1}{2} \langle T^{ij} \rangle \delta h_{ij} + \langle J^i \rangle \delta \Psi_i + \langle O_1 \rangle \delta \Phi_R \right).
\label{delta_S_ren_R}
\end{equation}
The above stress energy tensor $\langle T^{ij} \rangle$ contains finite contribution from the variation of $S_\mathrm{Dtr}$ with respect to $h_{ij}$, shifting the expressions of the Neumann theory (\ref{Ttt_Neumann})--(\ref{Tpsps_Neumann}). The expectation values in \eqref{delta_S_ren_R} are given by
\begin{align}
8 \pi G_N \langle T_{tt} \rangle &= - f_3 + 4 \phi_1 \phi_2 + \alpha \phi_1^2, \label{Ttt_Robin_F} \\
8 \pi G_N \langle T_{\theta\theta} \rangle &= - \frac{f_3}{2} -\alpha \phi_1^2, \label{Tthth_Robin_F} \\
8 \pi G_N \langle T_{\psi\psi} \rangle &= \sin^2 \theta \left(- \frac{f_3}{2} -\alpha \phi_1^2 \right), \label{Tpsps_Robin_F}\\
8 \pi G_N \langle J^t \rangle &= - a_1, \\
8 \pi G_N \langle O_1 \rangle &= \sqrt{2} \phi_1.
\end{align}

In our setup, we consider the Robin boundary conditions \eqref{Robin_BC} as the double trace deformation with vanishing source $\Phi_R=0$. 
From \eqref{Robin_BC}, we choose $\alpha=-\cot \zeta$, and the condition for the source is reduced to
\begin{equation}
\Phi_R = -\sqrt{2} \left( \phi_2 - \phi_1 \cot \zeta \right) = 0.
\label{source_robin}
\end{equation}
Under this condition,
the components of the stress energy tensor (\ref{Ttt_Robin_F})--(\ref{Tpsps_Robin_F}) become
\begin{align}
8 \pi G_N \langle T_{tt} \rangle &= - f_3 + 4 \phi_1 \phi_2 - \phi_1^2 \cot \zeta, \label{Ttt_Robin_Z} \\
8 \pi G_N \langle T_{\theta\theta} \rangle &= - \frac{f_3}{2} + \phi_1^2 \cot \zeta, \label{Tthth_Robin_Z} \\
8 \pi G_N \langle T_{\psi\psi} \rangle &= \sin^2 \theta \left(- \frac{f_3}{2} + \phi_1^2 \cot \zeta \right). \label{Tpsps_Robin_Z} 
\end{align}
When $\zeta=\pi/2$, these expressions reduce to those for the Neumann boundary conditions~(\ref{Ttt_Neumann})--(\ref{Tpsps_Neumann}). The total energy is given by
\begin{equation}
\mathcal{E}_R \equiv 8 \pi G_N \int \mathrm{d}\Omega_2 \langle T_{tt} \rangle = 4 \pi (- f_3 + 4 \phi_1 \phi_2 - \phi_1^2 \cot \zeta).
\end{equation}
This can be decomposed into individual contributions as
\begin{equation}
\mathcal{E}_R = \mathcal{E}_N + \mathcal{E}_\mathrm{Dtr} = \mathcal{E}_D + \mathcal{E}_\mathrm{LT} + \mathcal{E}_\mathrm{Dtr},
\label{total_energy_double}
\end{equation}
where $\mathcal{E}_\mathrm{Dtr}=-4 \pi \phi_1^2 \cot \zeta$ is the contribution of $S_\mathrm{Dtr}$. Among these, $\mathcal{E}_\mathrm{LT} + \mathcal{E}_\mathrm{Dtr}$ is interpreted as the energy stored on the AdS boundary.\footnote{See also \cite{allwright2016robin} for the relationship between the Robin boundary condition and the modification of the potential energy near a boundary, which may suggest that the parameter $\zeta$ for the Robin boundary condition controls the amount of the energy stored in the near-boundary region.}
Note that $\mathcal{E}_\mathrm{LT} + \mathcal{E}_\mathrm{Dtr}=0$ when $\zeta=\pi/2$, while it is not when $\zeta \neq \pi/2$ (and $\zeta \neq 0$, of course). The total charge and scalar expectation value are given by
\begin{equation}
\mathcal{Q} \equiv - 4 \pi a_1, \quad \langle \mathcal{O}_1 \rangle \equiv 4 \pi \sqrt{2} \, \phi_1.
\label{result_QO1_Robin}
\end{equation}

Using $\cot \zeta = \phi_2/\phi_1$, we can rewrite (\ref{Ttt_Robin_Z})--(\ref{Tpsps_Robin_Z}) as
\begin{align}
8 \pi G_N \langle T_{tt} \rangle &= - f_3 + 3 \phi_1 \phi_2,\\
8 \pi G_N \langle T_{\theta\theta} \rangle &= - \frac{f_3}{2} + \phi_1 \phi_2,\\
8 \pi G_N \langle T_{\psi\psi} \rangle &= \sin^2 \theta \left(- \frac{f_3}{2} + \phi_1 \phi_2 \right).
\end{align}
The total energy is expressed as
\begin{equation}
\mathcal{E}_R = 4 \pi (-f_3 +3 \phi_1 \phi_2).
\label{result_E_Robin}
\end{equation}
The trace of the energy momentum tensor can be written in the form
\begin{equation}
8 \pi G_N \langle T^i{}_i \rangle = - \phi_1 \phi_2 = - \cot \zeta \, \phi_1^2 = - \frac{\cot \zeta}{2} (8 \pi G_N)^2 \langle O_1 \rangle^2.
\end{equation}
This implies the spontaneous breaking of the conformal symmetry in the double trace deformed theory when the scalar operator acquires an expectation value.

The grand potential of the double trace deformed theory is also shifted from the Dirichlet and Neumann theories by a finite term as
\begin{equation}
\Omega_R = \Omega_N + \Omega_\mathrm{Dtr} = \Omega_D + \Omega_\mathrm{LT} + \Omega_\mathrm{Dtr},
\label{free_energy_double}
\end{equation}
where
\begin{equation}
\Omega_\mathrm{Dtr} = \mathcal{E}_\mathrm{Dtr} = - 4 \pi \cot \zeta \, \phi_1^2 = - 4 \pi \phi_1 \phi_2.
\label{free_energy_F_F}
\end{equation}
The expression of the grand potential in terms of the bulk integral is shifted from \eqref{free_energy_dirichlet} as
\begin{equation}
\Omega_R = 4 \pi \int_{r_h}^\infty \mathrm{d}r (1-e^{-\frac{\chi}{2}}) + 4 \pi \left( \frac{f_3}{2} - \phi_1 \phi_2 + r_h \right).
\label{free_energy_robin}
\end{equation}

\paragraph{RNAdS} For the RNAdS black holes (Eqs.~(\ref{metric_ansatz})--(\ref{RNAdSBG})), we have (the label of $D,N,R$ is removed because the scalar field is zero)
\begin{equation}
\mathcal{E} = 4 \pi r_h \left( 1 + r_h^2 + \frac{\mu^2}{2}\right).
\label{E_RN}
\end{equation}
The grand potential is
\begin{equation}
\Omega = 2 \pi r_h \left( 1-r_h^2 - \frac{\mu^2}{2} \right).
\label{F_RN}
\end{equation}
In thermal AdS, $r_h=0$, we obtain $\mathcal{E}=\Omega=0$. The Hawking-Page transition between the RNAdS and thermal AdS phases \eqref{RN_HP} occurs when the black hole reaches $\Omega=0$. The grand potential of the RNAdS \eqref{F_RN} is $\Omega > 0$ for $r_h < r_\mathrm{HP}$ and $\Omega <0$ for $r_h > r_\mathrm{HP}$ \eqref{RN_HP}. For the RNAdS, we can analytically check that \eqref{quant_stat_rel} is satisfied, where $\mathcal{Q} = 4 \pi Q = 4 \pi \mu r_h$ for the RNAdS.

\section{First law of thermodynamics}
\label{sec:1st}

To check numerical results, we wish to evaluate the first law of thermodynamics/black hole mechanics. If we regard solutions with nonzero $\phi_1$ and $\phi_2$ as the Dirichlet theory with explicit scalar source, the generalization of the first law of thermodynamics to the presence of nonzero scalar source and expectation values is given by\footnote{On general grounds, this first law in the presence of a nonzero scalar source follows from the fact that the grand potential is the generating function for responses of sources. In~\cite{Ishii:2018ucz}, this was discussed for the  holographic superconductor model same as this paper except in the probe limit with the planar AdS boundary. Recently in~\cite{Li:2020spf}, this scalar source contribution to the first law was derived by using Wald's formalism~\cite{Wald:1993nt,Iyer:1994ys}. See also~\cite{Lu:2013ura,Liu:2013gja,Lu:2014maa} for earlier discussions.}
\begin{equation}
\mathrm{d} \mathcal{E}_D = T_\mathrm{H} \mathrm{d} \mathcal{S}_\mathrm{BH} + \mu \mathrm{d}\mathcal{Q} - \langle \mathcal{O}_2 \rangle \mathrm{d} \Phi_D.
\label{1st_D}
\end{equation}
By the Legendre transform \eqref{Neumann_E}, this can be rewritten for the Neumann theory as
\begin{equation}
\mathrm{d} \mathcal{E}_N = T_\mathrm{H} \mathrm{d} \mathcal{S}_\mathrm{BH} + \mu \mathrm{d}\mathcal{Q} - \langle \mathcal{O}_1 \rangle \mathrm{d} \Phi_N.
\label{1st_N}
\end{equation}
Adding the double trace deformation \eqref{S_F_finite}, we can rewrite this for the double trace deformed theory. In this step, we can treat also $\alpha$
(defined by Eq.~(\ref{alphadef}))
as an independent variable.
By doing this, we can compare solutions with different values of $\alpha$.
We obtain
\begin{equation}
\mathrm{d} \mathcal{E}_R = T_\mathrm{H} \mathrm{d} \mathcal{S}_\mathrm{BH} + \mu \mathrm{d}\mathcal{Q} - \langle \mathcal{O}_1 \rangle \mathrm{d} \Phi_R - \frac{1}{8 \pi} \langle \mathcal{O}_1 \rangle^2 \mathrm{d} \alpha,
\label{1st_R}
\end{equation}
where $\mathrm{d} \Phi_R = -\sqrt{2} ( \mathrm{d} \phi_2 + \alpha \mathrm{d} \phi_1 + \phi_1 \mathrm{d} \alpha)$. The coefficient of the last term $1/(8 \pi)=1/(2 \cdot 4 \pi)$ is due to the normalization of $\langle \mathcal{O}_1 \rangle$~\eqref{result_QO1_Robin}. When we impose the Robin boundary conditions, i.e.~$\alpha=-\cot \zeta$ and $\Phi_R=0$ \eqref{source_robin}, this equation reduces to
\begin{equation}
\mathrm{d} \mathcal{E}_R = T_\mathrm{H} \mathrm{d} \mathcal{S}_\mathrm{BH} + \mu \mathrm{d}\mathcal{Q} + \frac{1}{8 \pi} \langle \mathcal{O}_1 \rangle^2 \mathrm{d} ( \cot \zeta).
\label{1st_R2}
\end{equation}
If we consider the family of solutions for fixed $\zeta$ (together with no source $\Phi_R=0$), the last term drops, and we obtain the first law of thermodynamics that contains the variation only of thermodynamic variables as
\begin{equation}
\mathrm{d} \mathcal{E}_R = T_\mathrm{H} \mathrm{d} \mathcal{S}_\mathrm{BH} + \mu \mathrm{d}\mathcal{Q}.
\end{equation}
However, if the last term is taken into account, we can use \eqref{1st_R2} as a relation useful to compare solutions where $\zeta$ varies in general. We can use any of the above equations to check numerical results because these are rewriting of the same relation.

\section{Comparison of entropy in microcanonical ensemble}
\label{sec:micro}

\begin{figure}[t]
\centering
\includegraphics[height=5cm]{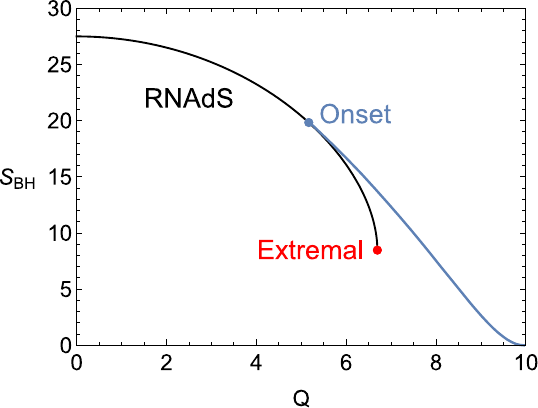}
\caption{Comparison of entropies in the microcanonical ensemble for $\mathcal{E}=10$, $\zeta/\pi=0.6$, and $q=1$.
The end point of the blue line is at $(\mathcal{S}_\mathrm{BH}, \mathcal{Q})=(0,9.955)$, and it corresponds to a charged Robin boson star.}
\label{fig:microS}
\end{figure}

In the main text, we have seen the phase structures in the grand canonical ensemble. We can also consider the microcanonical ensemble where the total energy (mass) $\mathcal{E}$ and charge $\mathcal{Q}$ are treated as independent variables.
In this ensemble, we can argue the fate of an unstable RNAdS black hole by comparing the entropies between solutions with and without scalar at the same $(\mathcal{E},\mathcal{Q})$ (see also Dirichlet boundary condition \cite{Maeda:2010hf,Basu:2010uz}).

In figure~\ref{fig:microS}, we show the entropies of the two kinds of the solutions in the $(\mathcal{Q},\mathcal{S}_\mathrm{BH})$ plane for $\mathcal{E}=10$, $\zeta/\pi=0.6$ and $q=1$. The black curve is the entropy of the RNAdS with $\mathcal{E}=10$. The extremal RNAdS is marked by the red dot, and the onset of instability for the branching of the hairy black holes is shown by the blue dot. When the RNAdS and hairy Robin black hole both exist at the same parameters $(\mathcal{E},\mathcal{Q},\zeta,q)$, the latter has the higher entropy than the former. 
We also examined other values of the parameters $(\mathcal{E},\mathcal{Q}, \zeta ,q)$ and found that hairy black holes have higher entropy than RNAdS when solutions overlap (see also the same comparison in the Dirichlet boundary condition \cite{Maeda:2010hf,Basu:2010uz}).
This implies that an unstable RNAdS can dynamically evolve into a hairy black hole in the microcanonical ensemble when it is perturbed and nonlinear time evolution is considered. In figure~\ref{fig:microS}, the zero entropy limit of the hairy Robin black hole is the zero size limit $r_h \to 0$ with diverging temperature $T_\mathrm{H} \to \infty$. The profile of the field variables $(f,\chi,\phi,A_t)$ approaches that of a charged Robin boson star.

\bibliography{bib_hairyrobin}

\end{document}